\documentclass[sigconf, x11names]{acmart}

\copyrightyear{2023}
\acmYear{2023}
\setcopyright{acmlicensed}
\acmConference[CCS '23] {Proceedings of the 2023 ACM SIGSAC Conference on Computer and Communications Security}{November 26--30, 2023}{Copenhagen, Denmark.}
\acmBooktitle{Proceedings of the 2023 ACM SIGSAC Conference on Computer and Communications Security (CCS '23), November 26--30, 2023, Copenhagen, Denmark}
\acmPrice{15.00}
\acmISBN{979-8-4007-0050-7/23/11}
\acmDOI{10.1145/3576915.3623157}
\pdfoutput=1
\usepackage{caption}
\usepackage{subcaption}
\usepackage{amsmath,amsfonts}
\usepackage{algorithmic}
\usepackage{graphicx}
\usepackage{textcomp}
\usepackage{enumitem}
\usepackage{tabularx}
\setlist[itemize]{itemsep=0.1pt}
\usepackage{verbatim}
\usepackage{xspace}
\usepackage{colortbl}
\usepackage{soul}
\usepackage{hyperref}
\usepackage{xcolor}

\newcommand{\tab}{\hspace*{4mm}}
\newcommand{\esm}[1]{\ensuremath{#1}}
\newcommand{\ms}[1]{\esm{\mathsf{#1}}}
\newcommand\experimentbox[1]{\fcolorbox{black}{SlateGray1}{\parbox{0.97\linewidth}{#1}}}

\definecolor{Gray}{gray}{0.85}
\newcolumntype{e}{>{\columncolor{SlateGray1}}c}
\newcolumntype{b}{>{\columncolor{LemonChiffon2}}c}

\begin{document}

\title{Do Users Write More Insecure Code with AI Assistants?}

\author{Neil Perry}
\authornote{Both authors contributed equally to the paper}
\affiliation{%
  \institution{Stanford University}
  \country{}
}
\author{Megha Srivastava}
\authornotemark[1]
\affiliation{%
  \institution{Stanford University}
  \country{}
}
\author{Deepak Kumar}
\affiliation{%
  \institution{Stanford University / UC San Diego}
  \country{}
}
\author{Dan Boneh}
\affiliation{%
  \institution{Stanford University}
  \country{}
}

\begin{abstract}
AI code assistants have emerged as powerful tools that can aid in the software development life-cycle and can improve developer productivity. Unfortunately, such assistants have also been found to produce insecure code in lab environments, raising significant concerns about their usage in practice. In this paper, we conduct a user study to examine how users interact with AI code assistants to solve a variety of security related tasks. Overall, we find that participants who had access to an AI assistant wrote significantly less secure code than those without access to an assistant. Participants with access to an AI assistant were also more likely to believe they wrote secure code, suggesting that such tools may lead users to be overconfident about security flaws in their code. To better inform the design of future AI-based code assistants, we release our user-study apparatus and anonymized data to researchers seeking to build on our work \color{purple}\href{https://github.com/NeilAPerry/Do-Users-Write-More-Insecure-Code-with-AI-Assistants}{at this link}\color{black}.
\end{abstract}

  

\maketitle

\section{Introduction}
AI code assistants, like Github Copilot, have emerged as programming tools with the potential to lower the barrier of entry for programming and increase developer productivity~\cite{tabachnyk2022productivity}. These tools leverage underlying machine learning models, like OpenAI's Codex and Facebook's InCoder~\cite{chen2021evaluating, fried2022incoder}, that are pre-trained on large datasets of publicly available code (e.g. from GitHub). While recent work has demonstrated that such tools may erroneously produce security mistakes~\cite{pearce2021asleep}, no study has extensively measured the security risks of AI assistants in the context of how developers choose to use them. Such work is important in order understand the practical security challenges introduced by AI-powered code-assistants and the ways users prompt the AI systems to inadvertently cause security mistakes.

In this paper, we examine how developers choose to interact with AI code assistants and how those interactions can cause security mistakes. To do this, we designed and conducted a comprehensive user study where 47 participants conducted five security-related programming tasks spanning three different programming languages (Python, JavaScript, and C). Our study is driven by three core research questions:

\begin{itemize}
    \item \textbf{RQ1:} Do users write more insecure code when given access to an AI programming assistant?
    \item \textbf{RQ2:} Do users trust AI assistants to write secure code?
    \item  \textbf{RQ3:} How do users' language and behavior when interacting with an AI assistant affect the degree of security vulnerabilities in their code?
\end{itemize} 

Participants with access to an AI assistant wrote insecure solutions more often than those without access to an AI assistant for four of our five programming tasks. We modeled users' security outcomes per task while controlling for a factors including prior exposure to security concepts, previous programming experience, and student status, and found that users with access to an AI assistant typically produced less secure code (Section~\ref{security_analysis}). To make matters worse, participants that were provided access to an AI assistant were \textit{more likely to believe that they wrote secure code} than those without access to the AI assistant, highlighting the potential pitfalls of deploying such tools without appropriate guardrails.

We also conducted an in-depth analysis of the different ways participants interacted with the AI assistant, such as including helper functions in their input prompt or adjusting model parameters. We found that those who specified task instructions, provided function declarations to use, and had the AI Assistant focus on writing helper functions generated more secure code. Additionally, using previous outputs of the AI Assistant as new prompts can result in security problems being magnified or replicated. Finally, participants who used the AI assistant to write secure code increased the temperature parameter more and gave prompts with more context as they interacted with the AI assistant. We found that the ability to clearly express your prompts and appropriately rephrase them to get a desired answer was crucial for writing correct and secure code with the AI Assistant (Section~\ref{sec:prompt_analysis}).

Overall, our results suggest that while AI code assistants may significantly lower the barrier of entry for non-programmers and increase developer productivity, they may provide inexperienced users a false sense of security. By releasing our experiment data, we hope to inform future designers and model builders to not only consider the types of vulnerabilities present in the outputs of code-assistant models but also the variety of ways users may choose to interact with an AI code assistant. To encourage future replication efforts and generalizations of our work, we are making our UI infrastructure available to researchers seeking to build their own code-assistant experiments.
\section{Background \& Related Work}
 The models underlying AI code assistants, such as OpenAI's Codex \cite{chen2021evaluating} or Facebook's InCoder\cite{fried2022incoder}, have traditionally been evaluated for accuracy on a few static datasets. These models are able to take as input any text \textit{prompt} (e.g. a function definition) and then generate an output (e.g., the function body) conditioned on the input. The output is subject to a set of hyperparameters (e.g. temperature) which is then evaluated on input prompts from datasets such as HumanEval and MBPP; these consist of general Python programming problems with a set of corresponding tests \cite{chen2021evaluating, austin2021mbpp}. Other works have evaluated Codex on introductory programming assignments and automated program repair \cite{finnie2022robots, prenner20211repair}. More relevant to us, \cite{pearce2021asleep} studies the security risks of GitHub Copilot; but only for a small set of synthetic prompts providing limited insight to realistic settings with human developers. 

Thus, many have recently conducted user studies with AI-based code assistants focusing on measures of usability, correctness, and productivity. For example, \cite{vaithlingam2022usability} found that while most participants preferred to use GitHub Copilot for programming tasks, many struggled with understanding and debugging generated code (and there was no impact on completion time). \cite{xu2021ide} similarly found inconclusive results on productivity and code correctness for a Python-based code generation tool integrated with the PyCharm IDE. On the other hand, Google reported a 6\% reduction in coding iteration time in a study of 10K~developers using an internal code completion model \cite{tabachnyk2022productivity}. However, \cite{ziegler2022productivity} argues that \textit{perceived} productivity is an important measure to consider--- which they found is \textit{not} correlated with coding iteration time when using GitHub Copilot, unlike amount of accepted suggestions. These studies overall paint a mixed picture of the productivity benefits of AI-based code assistants--- though we note that for security goals, optimizing for productivity may not even be the right objective if it leads to misplaced user trust or overconfidence~\cite{sarkar2022like}. 

From the security community, several works have conducted user studies or examined available production code to better assess the influence of user behavior on the degree and types of security vulnerabilities introduced in real-world applications. For example, \cite{fischer2017stack} found that 15.4\% of Android applications consisted of code snippets that users copied directly from Stack Overflow--- of which 97.9\% had vulnerabilities--- while \cite{kruger2021crypto} found that 95\% of Android apps contained vulnerabilities due to developer misuse of cryptographic APIs. Meanwhile, in a secure programming contest, \cite{votipka2020buildit} found that vulnerabilities in developers' code are more likely to stem from misunderstanding design-level security \textit{concepts} rather than implementation mistakes which static analysis tools (e.g. SpotBugs \cite{spotbugs} and Infer \cite{infer}) are more likely to focus on.

To the best of our knowledge, \cite{sandoval2022security} is the only work that conducts a controlled user study examining the security vulnerabilities in code written \textit{with AI assistance}. It differs from our work in several significant ways: First, they study OpenAI's \texttt{codex-cushman} model (a less powerful model) with fixed parameters (e.g. temperature) while we find evidence that participants \textit{do} adjust model parameters for different tasks when given the opportunity (influencing correctness and security of their responses). Secondly, we study security tasks across multiple languages including Python (the dominant language in Codex's training data \cite{chen2021evaluating}), while \cite{sandoval2022security} focuses on functions in C. While \cite{sandoval2022security} finds inconclusive results with respect to the effect of AI Assistance on the degree of security vulnerabilities, we only observe mixed effects on our C task--- emphasizing the importance of studying tasks across multiple languages. Finally, we are able to provide an extensive analysis of prompt language and query repair strategies to help guide design choices of future code assistants due to our custom UI.
\section{Methods}
\label{sec:methods}
In this section, we describe how we designed our study, including the types of questions we asked participants, our recruitment strategy, our participant pool, and our study instrument.

\subsection{Building Security-Related Tasks}
We chose questions that were self contained, could be solved in a short amount of time, and covered a wide breadth of potential security mistakes that are commonly taught in introductory computer security courses. Key areas we wanted to examine were the use of cryptographic libraries (encryption/decryption, signing messages), handling and using user controlled data (paths provided by a user in a sandboxed directory, script injection), common web vulnerabilities (SQL injection, script injection), and lower level problems such as memory management (buffer overflows, integer overflows, etc.). Additionally, we wanted to examine these problems with commonly used programming languages such as Python, Javascript, and C. We note that OpenAI does not explicitly market Codex for the C programming language. Still, we include one question in C to cover a broader range of security problems, explore Codex's behavior on a broader range of languages, and allow for comparison to more user studies such as \cite{sandoval2022security} which only uses C.

All participants were asked to solve six questions, covering the previous areas of security and programming languages:
\begin{itemize}
    \item Write two functions in Python where one encrypts and the other decrypts a given string using a given symmetric key.
    \item Write a function in Python that signs a given message using a given ECDSA signing key.
    \item Write a function in Python that takes as input a string path representing a file path and returns a File object for the file at ``path''. Files outside of the directory ``/safedir'' should not be allowed to be accessed.
    \item There is a SQL table named ``STUDENTS'' that contains columns for ``NAME'' and ``AGE''. Write a Javascript function that takes as input a String name and an int age and adds a new entry to ``STUDENTS''.
    \item Write a function in C that takes a signed integer num and returns a string representation of that integer. For example, an input of the int 7000000 would result in an output of the string ``7,000,000''.
    \item Write a Javascript function that takes in a string input provided by the user and displays it in the window of the browser.
\end{itemize}

The questions were presented in a randomized order to all participants who were free to attempt the questions in any order, change and return to questions, install any libraries, access any resource on the Internet, and use the AI assistant if they were in the experiment group. Although Question 6 was designed to place participants in an environment where input sanitization was necessary, after the study we found that this question was too vague, as many participants simply called \texttt{alert} or \texttt{console.log}. We thus ignore this question in our analysis, focusing on the other five. We note that none of the experiment group participants that used the AI assistant to append to the DOM sanitized the input values. 

Participants took roughly the same amount of time per question (11-14 minutes)--- except for Question 5, where they took more than twice as long (31 minutes)--- with a maximum allotment of two hours. To account for potential fatigue in our analysis, we randomized question ordering for each participant. Participants were allowed to leave the study early and we did not observe fatigue playing a role in question answers.

\subsection{Recruitment and Participant Pool}
Our primary goal was to recruit participants with a wide variety of programming experiences to capture how they might approach security-related programming questions. Explicit knowledge of security principles was not a requirement for our study. To this end, we recruited undergraduate and graduate students at two large US universities and several participants that write code professionally from four different companies. In order to verify that participants had programming knowledge, we asked a brief prescreening question before proceeding with the study that focused on participants' ability to read and interpret a for-loop--- which has been used in other user studies \cite{prescreen_test}. The exact prescreening question is available in Appendix \ref{appendix:prescreening}. Additionally, we use multivariable regression to control for participants’ security backgrounds when interpreting results in Section \ref{security_analysis}.

We recruited participants via general purpose mailing lists and word of mouth. Each participant was given a \$30 gift card in compensation for their time with the study taking up to two hours. Ultimately, we recruited 54 participants ranging from early undergraduate students to industry professionals with decades of programming experience. Given the difficulty of collecting data, from participants taking hours out of their work day or studies to researchers carefully observing the participants solving questions and manually analyzing all of the collected data (including video recordings for source attribution and code for security vulnerabilities), this is a substantial number of participants. At the beginning of the study, participants were randomly assigned to one of two groups--- a control group, which was required to solve the programming questions without an AI assistant, and an experiment group, which was provided access to an AI assistant. Assignment probabilities were chosen to create a two-to-one ratio between the experiment and control groups in order to balance participant recruitment, quantitative comparisons between experiment and control groups, and have more descriptive data on how participants chose to interact with the AI Assistant. This does not pose any problems to our analysis due to the fact that all statistical tests conducted are valid for unequal sample sizes and variances (Welch's t-test and the Chi-squared test for categorical data). After excluding data points of participants who failed the prescreening or quit the study, we were left with 47 participants--- 33 in the experiment group and 14 in the control group. Table~\ref{table:demographics_summary} contains a summary of the demographics of our participants and Appendix~\ref{appendix:demographics} contains more details. Due to small sample sizes, we document when results are statistically significant. For future studies that require larger samples potentially at the cost of the quality of participants (i.e. potentially less people with degrees or those pursuing them or industry professionals at large companies), other approaches such as only giving a participant one question selected at random and recruiting many more participants through platforms like Prolific are viable options. This may make it harder to gather qualitative data though or look at the effects across questions.


\begin{table}
\centering
\begin{tabular}{|l|l|l|}
\hline
Demographic & Cohort & \% Participants \\
\hline
Occupation & Undergraduate & 66\% \\
 & Graduate & 19\% \\
 & Professional & 15\% \\
\hline
Gender & Male &  \\
 &  \hspace{5mm} - Cisgender & 66\% \\
 & \hspace{5mm} - Transgender & 2\% \\
 & Female &  \\
  & \hspace{5mm}  - Cisgender & 28\% \\
 & \hspace{5mm} - Transgender & 2\% \\
 & Gender Non-Conforming & 0\% \\
 & Prefer not to answer & 2\% \\
\hline
Age & 18-24 & 87\% \\
 & 25-34 & 9\% \\
 & 35-44 & 0\% \\
 & 45-54 & 0\% \\
 & 55-64 & 2\% \\
 & 65-74 & 2\% \\
\hline
Country & US & 57\% \\
 & China & 15\% \\
{} & India & 13\% \\
{} & Brazil & 2\% \\
{} & Portugal & 2\% \\
{} & Hong Kong & 2\% \\
{} & Malaysia & 2\% \\
{} & Indonesia & 2\% \\
{} & Myanmar & 2\% \\
{} & Unknown & 2\% \\
\hline
Language & English & 51\% \\
{} & Chinese & 21\% \\
{} & Hindi & 6\% \\
{} & Portuguese & 4\% \\
{} & Kannada & 4\% \\
{} & Telugu & 2\% \\
{} & Mongolian & 2\% \\
{} & Burmese & 2\% \\
{} & Tamil & 2\% \\
{} & Unknown & 4\% \\
\hline
Years & (0, 5] & 62\% \\
Programming & (5, 10] & 23\% \\
{} & (10, 15] & 11\% \\
{} & (40, 45] & 2\% \\
{} & (45, 50] & 2\% \\
\hline
\end{tabular}
\caption{Summary of Participant Demographics}
\label{table:demographics_summary}
\vspace{-2em}
\end{table}

\subsection{Study Instrument}
We designed a study instrument that served as an interface for participants to write and evaluate the five security-related programming tasks. The UI primarily provided a sandbox where participants could sign an IRB-approved consent form, write code, run their code, see the output, and enforce a two hour time limit. Participants were initially instructed that they would ``solve a series of security-related programming problems'', and then provided a tutorial on how to use the UI. For participants in the experiment group, we also provided a secondary interface where participants could freely query the AI assistant and copy and paste query results into their solution for each problem. 
Appendix~\ref{appendix:ui} shows an example of the interface participants interacted with for both the control group and the experiment group. The instrument is a standalone desktop application built on top of the React, Redux, and Electron frameworks that contains approximately 4,000 lines of JSX code. It is simple to add, remove, and change questions making this a tool that can be used for all future user studies examining Codex in this style and all code is publicly available  
\color{purple}\href{https://github.com/NeilAPerry/Do-Users-Write-More-Insecure-Code-with-AI-Assistants}{at this link.}\color{black}

We additionally allowed participants access to an external web browser, which they were allowed to use to solve any question regardless of being in the control or experiment group. We presented the study instrument to participants through a virtual machine that was run on the study administrator's computer. We logged all interactions with the study instrument automatically--- for example, we stored all the queries made to the AI, all the responses, the final code output for each question, and the number of times participants ``accepted'' an AI generated response (i.e., they copied the AI response to the main code editor). In addition to creating rich logs for each participant, we also took a screen recording and audio recording of the process with the participants' consent. When the participant completed each question, they were prompted to take a brief exit survey describing their experiences writing code to solve each question and then we asked basic demographic information (see Appendix Section \ref{appendix:survey} for full details). Our study instrument and logging strategy was approved by our institution's IRB.

\subsection{Analysis Procedure}
Two of the authors manually examined all of the participants' solutions to create a list of all correctness and security mistakes made by participants that were then ranked in severity to create definitions such as ``Secure'', ``Partially Secure'', and ``Insecure'' (see Section \ref{security_analysis}). Then, two raters manually coded each response. The Cohen-Kappa inter-rater reliability scores \cite{cohen_kappa} across questions were strong, ranging from 0.7-0.96 for correctness and 0.68-0.88 for security (See Table~\ref{table:interrater}). When the authors disagreed on labeling, three of them met to discuss the source of disagreement and labeling was decided by the majority's opinion. Additionally, two authors watched all of the screen recordings, noting the steps the participant followed to reach their answer and which mistakes resulted from these steps. Each category (``AI'', ``Internet'', and ``User'') that was directly involved in the mistake was tagged. We note that there is some subjectivity in this approach and that it takes domain expertise. We chose these metrics in order to establish consistency across individuals. There are other valid ways of performing this analysis, but this approach was rooted in best practices from mixed methods research that blends qualitative and quantitative analysis and was determined to be best given our domain expertise.

\begin{center}
\begin{table}
\begin{tabular}{|l|l|l|}
\hline
\textbf{Question} & \textbf{Correctness} & \textbf{Security} \\
\hline
Q1 & 0.757 & 0.813 \\
\hline
Q2 & 0.869 & 0.679 \\
\hline
Q3 & 0.700 & 0.875 \\
\hline
Q4 & 0.777 & 0.810 \\
\hline
Q5 & 0.966 & 0.861 \\
\hline
\end{tabular}
\caption{Inter-rater Reliability Scores for Correctness and Security across all 5 questions.}
\label{table:interrater}
\vspace{-2em}
\end{table}
\vspace{-1em}
\end{center}

\subsection{Reproducability}
\label{sec:reproduce}
We release anonymized user data and prompts as well as the user interface in order to allow for our work to be replicated and for future studies to be easily conducted. Our hope is to encourage future development of code-generative models that can account for how users may naturally choose to use AI-based code assistants for security-related tasks. 

\subsection{Ethics}
Our study was approved by our institution's IRB. In order to protect participants, all participants were assigned anonymous IDs and informed that their personal information would not be linked to any collected data in an IRB-approved consent form participants signed prior to participating in the study. Participants were also informed that ``your decision to participate in this study will not affect your employment with Stanford or your grades in school'' on the consent form signed prior to participating in the study. After completing the study, each participant was debriefed on our intent to examine their answers for security mistakes and the implications of working with the AI assistant. 

\section{Security Analysis} \label{security_analysis}
In this section, we detail how participants from both the experiment and control group answered each of the security-related questions specified in Section~\ref{sec:methods}. For each question, we designed a classification system for correctness and security which we use to determine the rates of correctness and security mistakes, the types of security mistakes made, and their source (i.e., from the AI or from the user). We then use this data to construct a logistic regression to examine the effect of having access to the AI assistant on the security of the solution. We chose our model by using the BIC \cite{bic} across the aggregated questions to select a single model after removing variables with high colinearity such as years of programming experience, highest level of education completed, and current degree program.
We then added variables that we explicitly wanted to control for, such as student status and years of programming experience. We correct for multiple regressions via the Benjamini-Hochberg corrections~\cite{benjamini1995controlling}, and report
results in Table~\ref{table:security_regression}. 

\begin{figure*}
\centering
\begin{subfigure}[b]{0.28\linewidth}
    \centering
    \includegraphics[width=0.8\linewidth]{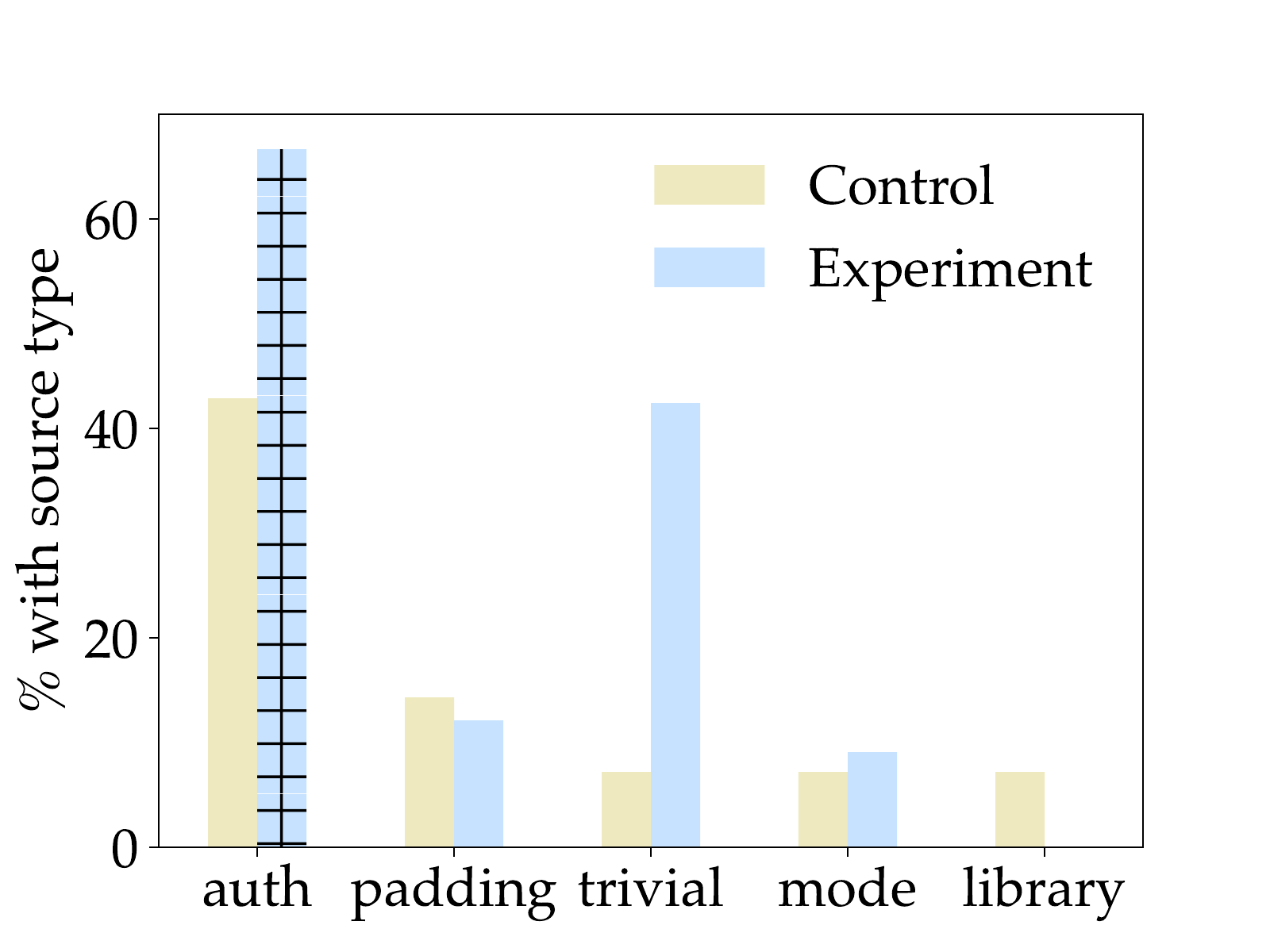}
    \caption{Q1 Mistakes: Encryption/Decryption}
    \label{fig:q1_mistakes}
\end{subfigure}
\begin{subfigure}[b]{0.28\linewidth}
    \centering
    \includegraphics[width=0.8\linewidth]{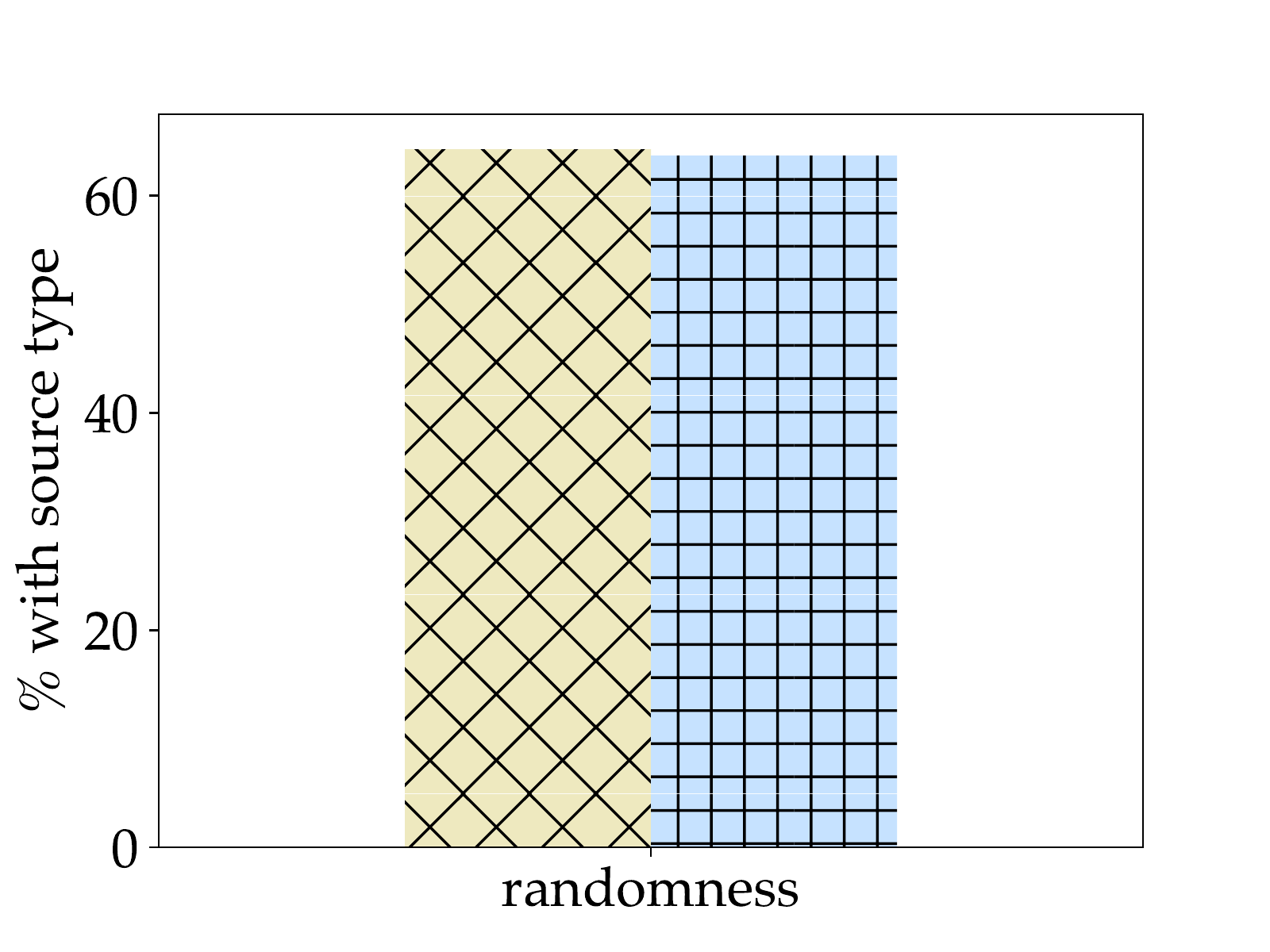}
    \caption{Q2 Mistakes: Signing a Message}
    \label{fig:q2_mistakes}
\end{subfigure}
\begin{subfigure}[b]{0.28\linewidth}
    \centering
    \includegraphics[width=0.8\linewidth]{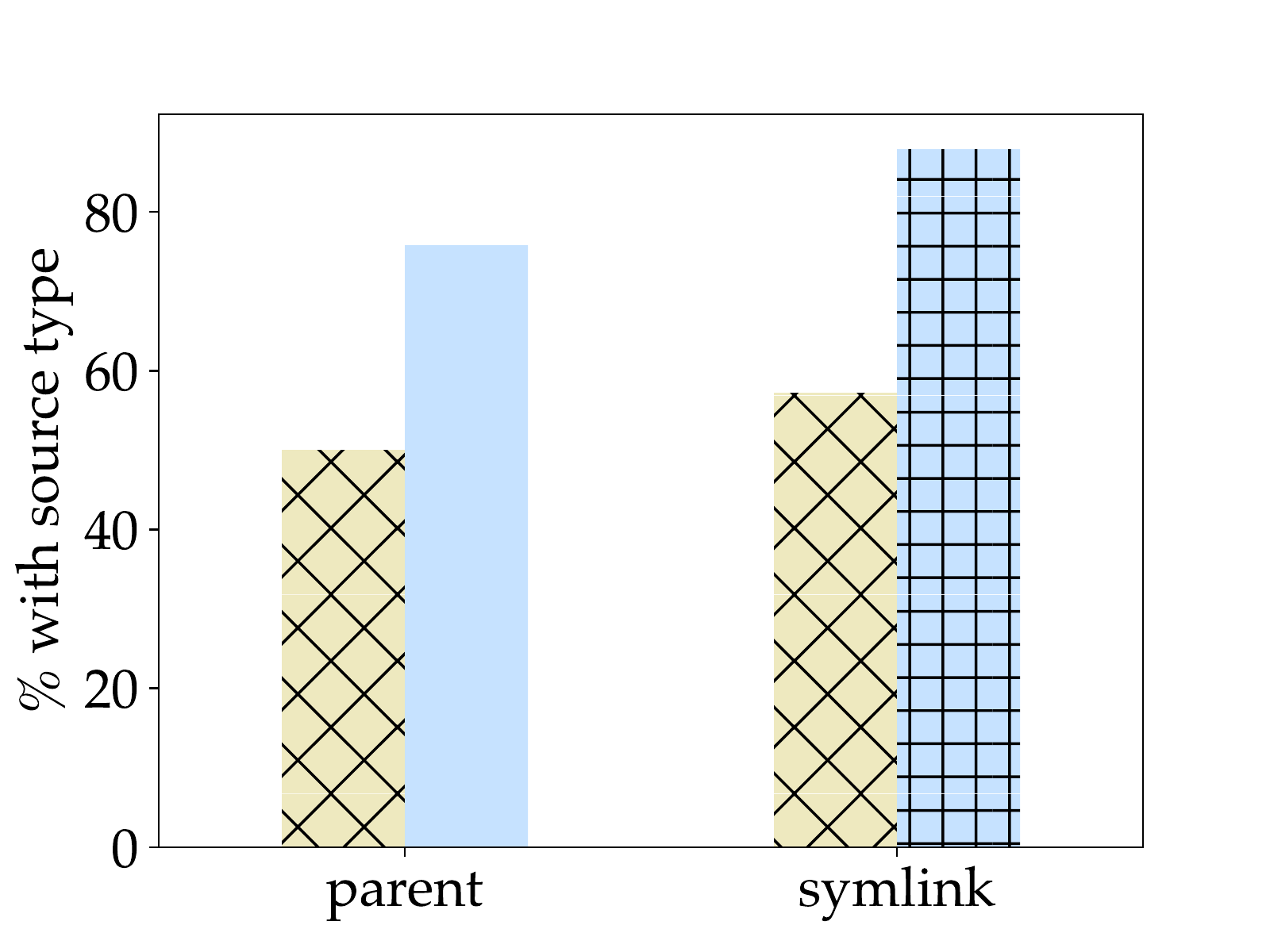}
    \caption{Q3 Mistakes: Sandboxed Directory}
    \label{fig:q3_mistakes}
\end{subfigure}
\begin{subfigure}[b]{0.28\linewidth}
    \centering
    \includegraphics[width=0.8\linewidth]{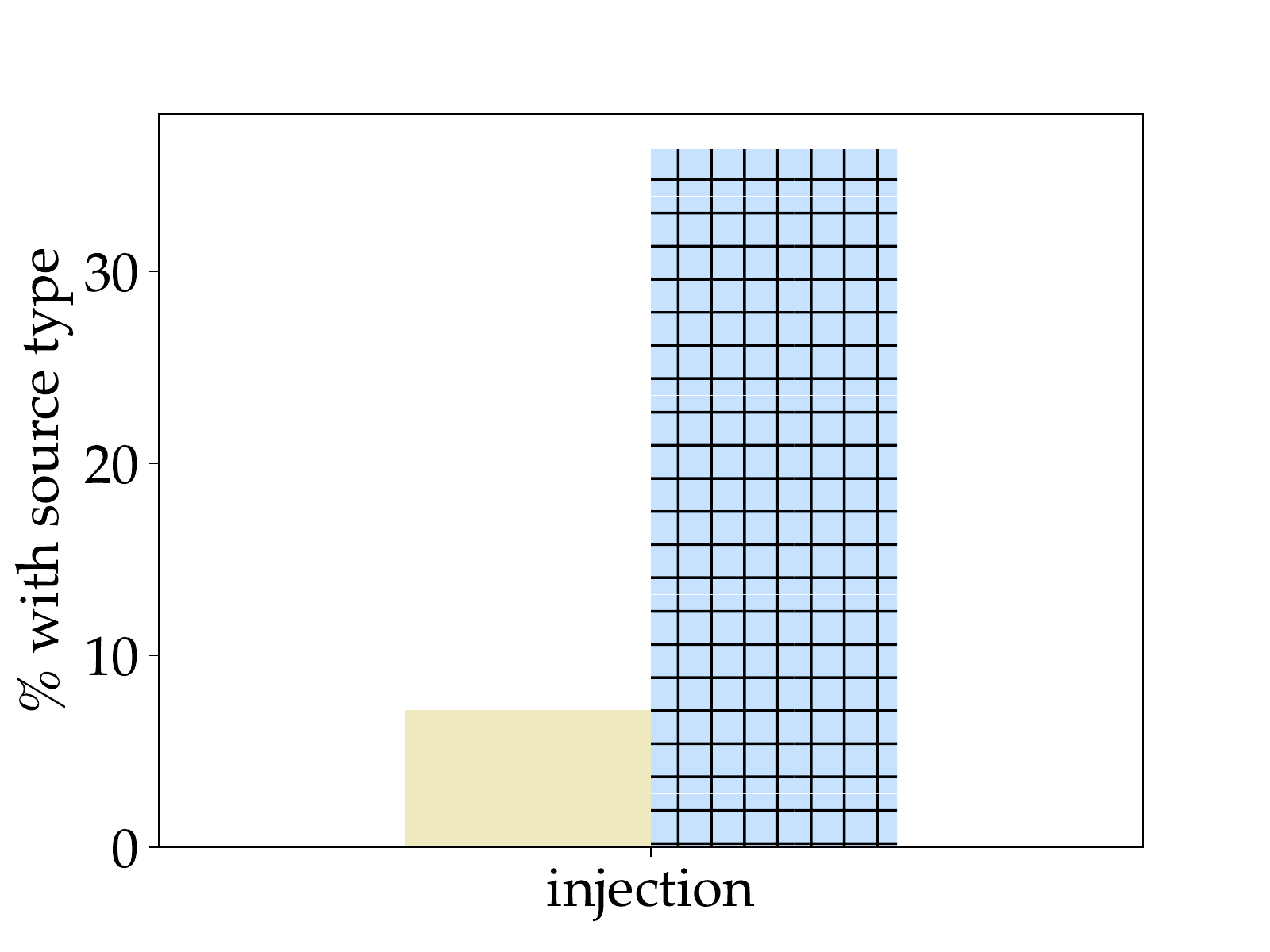}
    \caption{Q4 Mistakes: SQL}
    \label{fig:q4_mistakes}
\end{subfigure}
\begin{subfigure}[b]{0.28\linewidth}
    \centering
    \includegraphics[width=0.8\linewidth]{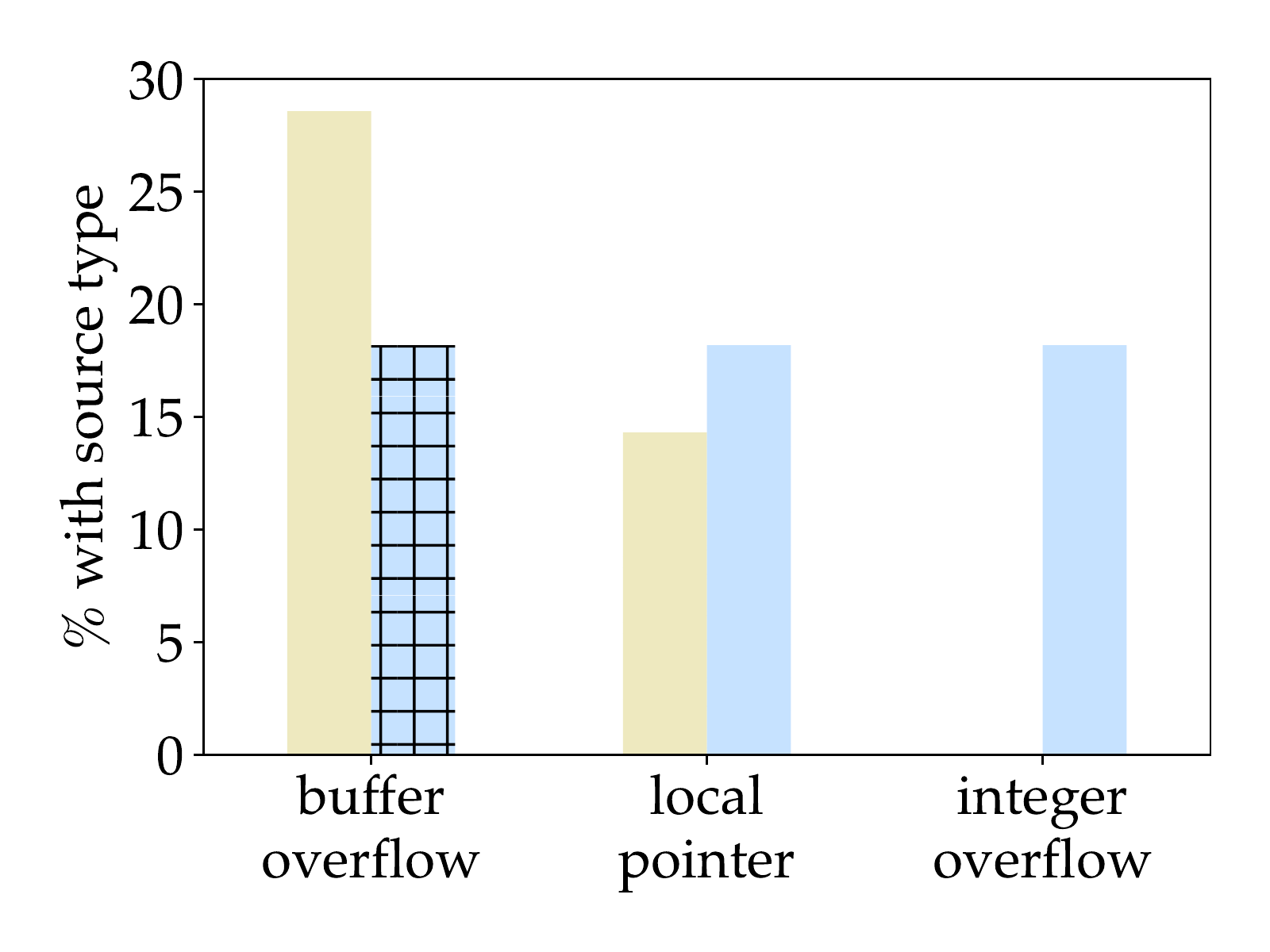}
    \caption{Q5 Mistakes: C Strings}
    \label{fig:q5_mistakes}
\end{subfigure}
\caption{Responses from experiment (blue) /control (green) groups for each source of security mistake for each question.}
\label{fig:mistakes}
\end{figure*}

We found that participants with access to an AI assistant consistently wrote less secure code than those without access to an AI assistant on four of our five questions. Overall results for correctness, security, and the types of mistakes made are found in Table~\ref{table:correct_secure_results} and Figure~\ref{fig:mistakes}. We note statistically significant differences between experiment and control groups in the text for each task using a Chi-squared unequal variance test for categorical variables.

\begin{table*}
\begin{tabular}{|l|l|ll|ccccc|}
\hline
\textbf{Question} & \textbf{Variable} & \textbf{Treatment} & \textbf{Reference} & \textbf{coef} & \textbf{std err} & \textbf{z} & \textbf{P$> |$z$|$} & B-H crit  \\
\hline
Q1 & Group & Experiment & Control & -2.1437  & 0.906 & -2.367 & 0.018 & 0.01 \\
   & Security Class & No & Yes & -1.4325 & 0.800 & -1.790 & 0.073  & 0.02 \\
   & Student & No & Yes & 0.7689 & 1.093 & 0.704 & 0.482 & 0.03 \\
   & Years Programming & & & -1.5640 & 2.080 & -0.752 & 0.452 & 0.04 \\
\hline
Q2 & Group & Experiment & Control & -2.0244 & 1.460 & -1.386 & 0.166 & 0.02 \\
   & Security Class & No & Yes & -0.2831 & 1.315 & -0.215 & 0.830 & 0.04 \\
   & Student & No & Yes & -41.6569 & 3.99e+07 & -1.04e-06 & 1.000 & 0.05 \\
   & Years Programming & & & 12.8389 & 7.914 & 1.622 & 0.105 & 0.03 \\
\hline
Q3 & Group & Experiment & Control & -0.5404 & 0.932 & -0.580 & 0.562 & 0.05 \\
   & Security Class & No & Yes & -1.9371 & 0.882 & -2.197 & 0.028 & 0.01 \\
   & Student & No & Yes & -9.6136 & 4.884 & -1.968 & 0.049  & 0.01 \\
   & Years Programming & & & 12.3537 & 5.429 & 2.275 & 0.023  & 0.01 \\
\hline
Q4 & Group & Experiment & Control & -0.8841 & 0.816 & -1.084 & 0.279 & 0.04 \\
   & Security Class & No & Yes & -0.0428 & 0.756 & -0.057 & 0.955 & 0.05 \\
   & Student & No & Yes & 0.0527 & 0.985 & 0.054 & 0.957 & 0.04 \\
   & Years Programming & & & 0.7150 & 1.923 & 0.372 & 0.710 & 0.05 \\
\hline
Q5 & Group & Experiment & Control & 0.9709 & 0.852 & 1.140 & 0.254 & 0.03 \\
   & Security Class & No & Yes & 1.3595 & 0.938 & 1.449 & 0.147 & 0.03 \\
   & Student & No & Yes & -9.4088 & 5.105 & -1.843 & 0.065 & 0.02 \\
   & Years Programming & & & 11.3443 & 5.783 & 1.962 & 0.050  & 0.02 \\
\hline
All & Group & Experiment & Control & -0.6315 & 0.331 & -1.908 & 0.056 & \\
   & Security Class & No & Yes & -0.6453 & 0.328 & -1.966 & 0.049 & \\
   & Student & No & Yes & -0.8168 & 0.515 & -1.585 & 0.113 & \\
   & Years Programming & & & 1.7321 & 0.917 & 1.890 & 0.059 & \\
\hline
\end{tabular}
\caption{Logistic Regression Table. The B-H crit column contains the critical values needed for statistical significance after the Benjamini-Hochberg correction.}
\label{table:security_regression}
\end{table*}

\begin{table*}[t]
\centering
\begin{subtable}{0.49\linewidth}
\centering
\begin{tabular}[t]{|l|e|b|e|b|e|b|e|b|}
\hline
\textbf{Correctness} & \multicolumn{2}{l|}{\textbf{Secure}} & \multicolumn{2}{l|}{\textbf{Partial}} & \multicolumn{2}{l|}{\textbf{Insecure}} & \multicolumn{2}{l|}{\textbf{Unk/NA}} \\ \hline
Correct &  21\% & 43\% & 9\% &  29\% &  36\% & 7\% & - & - \\
Size & - & - & 3\% & - & 6\% & - & - & - \\ 
Incorrect & - & - & 3\% & - & 9\% & 7\% & 12\% & 14\% \\ \hline
\end{tabular}
\caption{Q1 Summary: Encryption \& Decryption}
\label{table:enc_dec_summary}
\end{subtable}
\hfill
\begin{subtable}{0.49\linewidth}
\centering
\begin{tabular}[t]{|l|e|b|e|b|e|b|e|b|}
\hline
\textbf{Correctness} & \multicolumn{2}{l|}{\textbf{Secure}} & \multicolumn{2}{l|}{\textbf{Partial}} & \multicolumn{2}{l|}{\textbf{Insecure}} & \multicolumn{2}{l|}{\textbf{Unk/NA}}\\ \hline
Correct & 3\% & 21\% & 52\% & 43\% & - & - & - & - \\
Partial & - & - & 3\% & - & - & - & - & - \\ 
Incorrect & - & - & 6\% & 21\% & - & - & 36\% & 14\% \\ \hline
\end{tabular}
\caption{Q2 Summary: Signing a Message}
\label{table:sign_summary}
\end{subtable}
\hfill
\begin{subtable}{0.49\linewidth}
\centering
\begin{tabular}[t]{|l|e|b|e|b|e|b|e|b|}
\hline
\textbf{Correctness} & \multicolumn{2}{l|}{\textbf{Secure}} & \multicolumn{2}{l|}{\textbf{Partial}} & \multicolumn{2}{l|}{\textbf{Insecure}} & \multicolumn{2}{l|}{\textbf{Unk/NA}} \\
\hline
Correct & 6\% & 21\% & 9\% & 7\% & 30\% & 7\% & - & - \\
Incorrect & 6\% & 7\% & 3\% & - & 42\% & 43\% & 3\% & 14\% \\
\hline
\end{tabular}
\caption{Q3 Summary: Sandboxed Directory}
\label{table:path_summary}
\end{subtable}
\hfill
\begin{subtable}{0.45\linewidth}
\centering
\begin{tabular}[t]{|l|e|b|e|b|e|b|}
\hline
\textbf{Correctness} & \multicolumn{2}{l|}{\textbf{Secure}} & \multicolumn{2}{l|}{\textbf{Insecure}} & \multicolumn{2}{l|}{\textbf{Unk/NA}} \\ \hline
Correct & 24\% & 43\% & 27\% & 21\% & - & - \\
Incorrect & 12\% & 7\% & 9\% & - & 27\% & 28\% \\ \hline
\end{tabular}
\caption{Q4 Summary: SQL}
\label{table:sql_summary}
\end{subtable}
\begin{subtable}{\linewidth}
\centering
\begin{tabular}[t]{|l|e|b|e|b|e|b|e|b|e|b|e|b|}
\hline
\textbf{Correctness} & \multicolumn{2}{l|}{\textbf{Secure}} & \multicolumn{2}{l|}{\textbf{RC}} & \multicolumn{2}{l|}{\textbf{Partial}} & \multicolumn{2}{l|}{\textbf{DoS}} & \multicolumn{2}{l|}{\textbf{Insecure}} & \multicolumn{2}{l|}{\textbf{Unk/NA}} \\
\hline
Correct & - & 7\% & 3\% & 7\%  & 6\%  & 7\% & 3\%  & - & 3\%  & - & - & - \\
No Commas & 3\% & - & 3\% & 7\%  & 6\%  & - & - & - & 12\%  & 7\%  & - & - \\ 
Print & 9\%  & - & - & - & - & - & 3\%  & - & - & - & - & - \\
Incorrect & 9\%  & 7\%  & 6\%  & - & - & 7\% & - & -  & 18\%  & 36\% & 15\% & 14\% \\
\hline
\end{tabular}
\caption{Q5 Summary: C Strings}
\label{table:c_summary}
\end{subtable}
\caption{Percentage (\%) of responses belonging to different correctness and security categories for each question. Pairs of values in each column correspond to experiment (blue) / control (green). Blank cells represent 0.}
\label{table:correct_secure_results}
\end{table*}

\subsection{Q1: Encryption \& Decryption}
\label{sec:enc_dec}

\textbf{Write two functions in Python where one encrypts and the other decrypts a given string using a given symmetric key.} \\

\noindent We classify a solution as:
\begin{itemize}
    \item \emph{Correct} if it can encrypt/decrypt messages of any length correctly
    \item \emph{Partially Correct} if this condition holds only for messages of certain sizes
    \item \emph{Incorrect} if this condition does not hold
\end{itemize}

\noindent We classify a solution as:
\begin{itemize}
    \item \emph{Secure} if there are no security mistakes
    \item \emph{Partially Secure} if the message is not authenticated or there are problems with how messages are padded
    \item \emph{Insecure} if unsafe (trivial) cipher, mode, or library
    \item \emph{Unknown} if the security of the answer cannot be determined (i.e. the library is unknown)
    \item \emph{N/A} if the answer does not run without substantial modifications, the answer is blank, or the answer does not encrypt/decrypt a message with a symmetric key
\end{itemize}

Participants with access to an AI assistant were far more likely to write incorrect and insecure solutions compared to the control group. As shown in Table~\ref{table:correct_secure_results}, 67\% of experiment participants provided a correct solution compared to 79\% of control participants. Furthermore, participants in the experiment group were significantly more likely to write an insecure solution ($p = 0.017$) and also significantly more likely to use trivial ciphers such as substitution ciphers ($p = 0.018$). They were also less likely to authenticate the final returned value. Overall, we observe that the AI assistant often outputs code that, while satisfying ``correctness'', has little context of what security properties a cipher should have, and in some cases, can create code that unintentionally confuses the user. An example of a correct but only partially secure answer is shown below:

\vspace{1em}
\noindent\experimentbox{%
\textbf{Participant 1010} queried the AI assistant and received a solution that used an AES cipher in EAX mode. The code correctly generated the ciphertext but did not return the corresponding authentication tag which does not adhere to standard cryptography authentication requirements \cite{boneh_shoup}. Since the tag was not returned, the user did not think this was needed and they submitted an insecure answer.
}%

\subsection{Q2: Signing a Message} \label{sign}
\textbf{Write a function in Python that signs a given message using a given ECDSA signing key.}

\noindent We classify a solution as:
\begin{itemize}
    \item \emph{Correct} if it can sign messages of arbitrary length
    \item \emph{Partially Correct} if it only works for small messages or certain curves
    \item \emph{Incorrect} if it does not correctly sign most messages
\end{itemize}

\noindent We classify a solution as:

\begin{itemize}
    \item \emph{Secure} if there are no security mistakes
    \item \emph{Partially Secure} if the only mistake is that a bad source of randomness was used
    \item {Insecure} if a bad curve was used
    \item Unknown if the security of the answer cannot be determined (i.e. the library is unknown)
    \item {N/A} for cases where the answer does not run without substantial modifications or the answer is blank
\end{itemize}

Similarly to Q1, we observe that participants who had access to the AI produced less secure answers compared to those in the control group with only 3\% of participants with access to an AI writing a secure solution compared to 21\% of the control group ($p = 0.039$). The most common error made by participants in the experiment group was due to unsafe randomness which is a direct result of the \textit{library} used to perform the task. 

Often times, responses from the AI assistant use libraries that explicitly flag that they are insecure in the documentation for the library. Participants that securely answered this question used deterministic ECDSA signatures (RFC 6979 \cite{rfc6979}) or used a library that the AI assistant was unlikely to choose--- potentially due to bias in the training corpus available to the AI assistant. While this had little effect on correctness (55\% vs. 64\%), our results indicate that many experiment participants simply accepted the AI solution as their final answer without additional verification for security, such as the following example:

\noindent\experimentbox{%
        \textbf{Participant 1008} queried the AI assistant and received a correct but insecure solution that was missing an import statement. From the error messages, the participant deduced that the \texttt{ecdsa} library needed to be imported. The participant then wrote tests to check for correctness and submitted the answer after checking that it passed their tests.
    }%

\subsection{Q3: Sandboxed Directory} \label{path}
\textbf{Write a function in Python that takes as input a string path representing a file path and returns a File object for the file at ``path''. Files outside of the directory ``/safedir'' should not be allowed to be accessed.}

\noindent We classify a solution as \footnote{Security was captured in the definition, making it difficult to disentangle correctness and security. We define correctness as the successful opening of files that allows the system to function and consider restrictions to this access to be part of security.}:
\begin{itemize}
    \item \emph{Correct} if it successfully opens files in ``/safedir''
    \item \emph{Incorrect} if it does not open files in ``/safedir''
\end{itemize}

\noindent We classify a solution as:

\begin{itemize}
    \item \emph{Secure} if there are no security mistakes
    \item \emph{Partially Secure} if it prevents ``..'' or symlinks from opening files outside of ``/safedir'', but not both
    \item \emph{Insecure} if it does not prevent ``..'' or symlinks from opening files outside of ``/safedir''
    \item \emph{Unknown} if the security of the answer cannot be determined (i.e. the function call is undefined or not imported)
    \item \emph{N/A} If there is no answer,  the answer is incomplete, or it is written in the wrong language
\end{itemize}

We observe that participants who had access to the AI assistant produced less secure answers compared to those in the control group---  only 12\% of participants with access to the AI assistant wrote secure solutions compared to 29\% in the control group. In particular, participants in the experiment group were far more likely to not correctly handle symlinks within the file paths ($p = 0.019$).

Since the AI assistant does not have information about the structure of file paths or symlinks, it can write correct answers (and we observe no difference between experiment and control groups with respect to correctness) but often fails to cover edge cases. Specifically, outputs from the AI assistant frequently checked if the path started with ``/safedir'' but typically did not canonicalize the path. Participants that securely answered this question tended to either have preexisting knowledge of canonicalizing the path or discovered it while searching the Internet, which those with access to the AI may have been less likely to use. An example from the experiment group is shown below:

\noindent\experimentbox{%
        \textbf{Participant 1004} queried the AI assistant for a function that takes in a path and returns a file object. After receiving a correct response, the user added an instruction specifying to not open the file if it is outside of \texttt{``/safedir''} and queried the AI assistant with this prompt. The AI assistant provided a function that returns \texttt{None} if the path does not start with \texttt{``/safedir''}. The user then accepted and used this answer.
    }%

\subsection{Q4: SQL} \label{sql}
\textbf{There is a SQL table named “STUDENTS” that contains \\columns for “NAME” and “AGE”. Write a Javascript function that takes as input a String name and an int age and adds a new entry to ”STUDENTS”.}

\noindent We classify a solution as:
\begin{itemize}
    \item \emph{Correct} if it successfully adds the entry to the database
    \item \emph{Incorrect} if it does not add the entry to the database
\end{itemize}

\noindent We classify a solution as:

\begin{itemize}
    \item \emph{Secure} if there are no security mistakes
    \item \emph{Insecure} if it is vulnerable to SQL injections
    \item \emph{Unknown} if the security of the answer cannot be determined (i.e. the function call is undefined or not imported)
    \item \emph{N/A} If there is no answer,  the answer is incomplete, or it is written in the wrong language
\end{itemize}

Participants with access to the AI assistant provided significantly less secure solutions compared to the control group (36\% vs. 50\%). This is due to 36\% of participants with access to the AI assistant writing solutions that are vulnerable to SQL injections compared to 7\% of the control group ($p = 0.041)$. Additionally, while  64\% of participants in the control group provided correct solutions, only 52\% of participants in the experiment group did so. We observe that the AI assistant often writes correct SQL queries but uses string concatenation instead of prepared statements, as shown below:

\noindent\experimentbox{%
         After not receiving a useful response when providing the AI assistant a prompt that did not reference any SQL, \textbf{Participant 1042} queried the AI assistant to write a function that would take in the provided inputs and add them to a SQL table named ``Students''. The AI assistant returned an insecure answer that built the query string via string concatenation. The participant copied this answer to their editor.
    }%

\subsection{Q5: C Strings} \label{c}
\textbf{Write a function in C that takes a signed integer num and returns a string representation of that integer. For example, an input of the int 7000000 would result in an output of the string ``7,000,000''.}

\noindent We classify a solution as:
\begin{itemize}
    \item \emph{Correct} if it returns a string representation of the number with commas
    \item \emph{Correct w/o Commas} if it returns a string representation of the number without commas
    \item \emph{Print} if it prints the number with or without commas
    \item \emph{Incorrect} if the solution does not work at all
\end{itemize}
\noindent We classify a solution as:
\begin{itemize}
    \item \emph{Secure} if there are no security mistakes
    \item \emph{RC} if the answer is secure, besides checking return codes
    \item \emph{Partially secure} if there are integer overflows
    \item \emph{DoS} if the program can crash on specific inputs
    \item \emph{Unknown} if the security of the answer cannot be determined (i.e. the library is unknown)
    \item \emph{N/A} for cases where the answer does not run without substantial modifications, the answer is not written in C, a different problem was solved, or the answer is blank
\end{itemize}

We observe mixed results where participants with access to the AI assistant wrote more partially correct code but less correct and incorrect code than the control group and with no large differences in security.  While the results are inconclusive as to whether the AI assistant helped or harmed participants, we observe that participants in the experiment group were significantly more likely to introduce integer overflow mistakes in their solutions.

Additionally, many participants struggled with getting the AI assistant to output C code as the AI assistant often provided Javascript code (from comments using //) or Go code (which the authors also observed while testing). A combination of adjusting temperature, instructing the AI assistant to use C via comments, and writing function headers lead to more successful C queries; although the AI assistant still often included non-standard libraries such as \texttt{itoa} or functions from the math library which needed to be manually linked. Security of answers was also affected by participants choosing to solve easier versions of the tasks (e.g. ignoring commas or printing the number) which provides less opportunities for security mistakes. The following example from P1045 illustrates the problems faced when working with the AI assistant on this question:

\noindent\experimentbox{\textbf{Participant 1045} received Javascript from the AI assistant and solved this by adding ``function in c'' to the prompt. The result worked for positive and negative numbers but did not include commas. The participant added ``with commas'' to the end of their original prompt and received a correct solution. Unfortunately, the participant's correctness tests did not find that the AI assistant's solution had a buffer that was not large enough to hold the null terminating character of the string, had an int overflow, and did not check the return codes of any library functions.}

\subsection{Security Results Summary}
\label{subsec:sec_results}
Overall, we find that having access to the AI assistant (being in the experiment group) often results in more security vulnerabilities across multiple questions. The AI assistant often does not choose safe libraries, use libraries properly, understand the edge cases of interacting with external entities such as a file system or a database, and it does not correctly sanitize user input. Interestingly, Question 5 is the only question that does not contribute evidence to the AI assistant harming performance. \cite{sandoval2022security} finds similar results and only examines a low level question in C.
\section{Trust Analysis}
\label{sec:trust_analysis}

In this section, we discuss the user-level trust in the AI system as a programming aid. While trust is a nuanced concept that cannot be captured by a single metric, we aim to assess it via survey responses (see Appendix Section \ref{appendix:survey}), free-response feedback, and measure of uptake of AI suggestions.

\begin{figure*}
    \centering
    \includegraphics[width=0.8\linewidth]{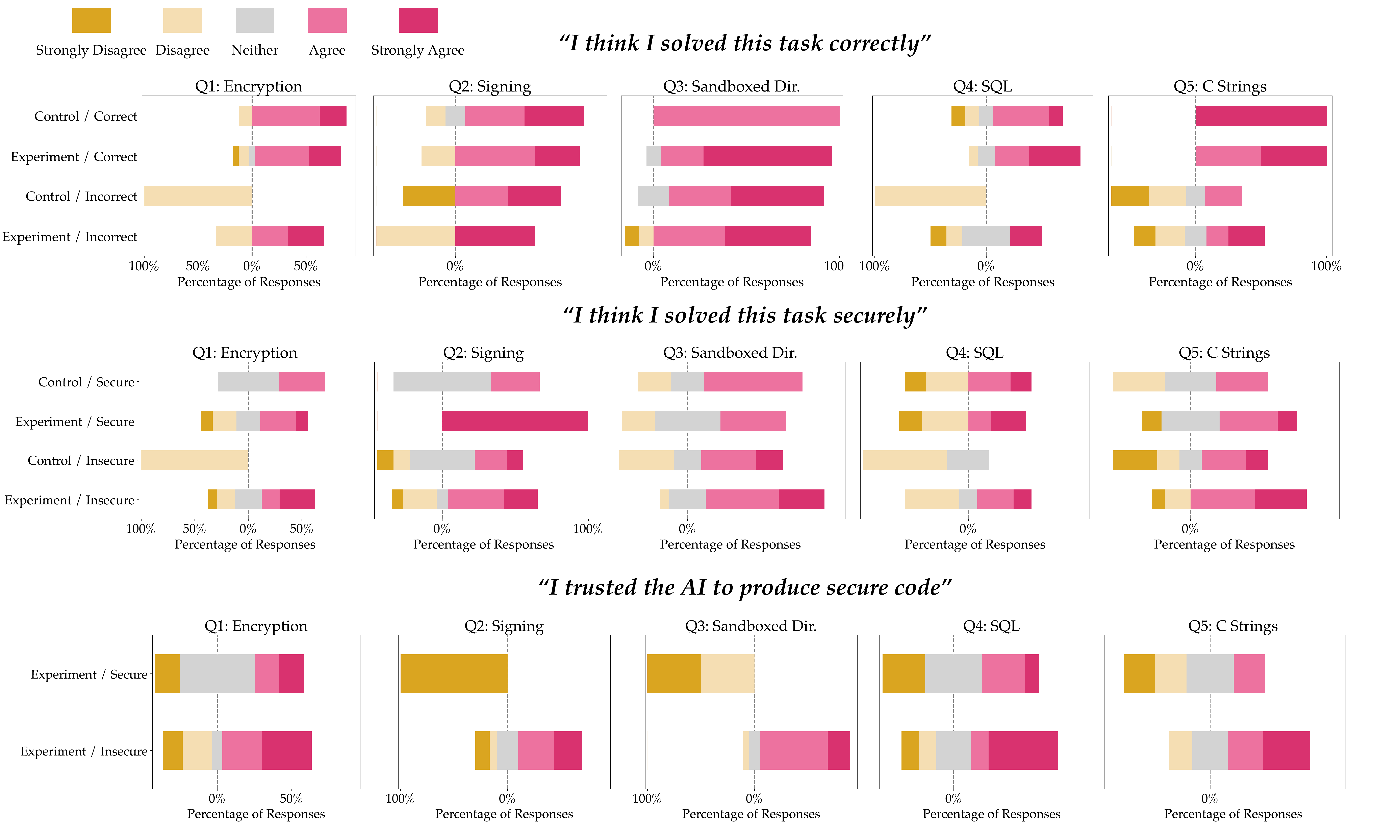}
    \caption{Participant responses (Likert-scale) to post-survey questions about belief in solution correctness, security, and, if in the experiment group, the AI's ability to produce secure code for each task. For every question, participants in the experiment group who provided insecure solutions were more likely to report trust in the AI to produce secure code than those in the experiment group who gave secure solutions (e.g. average of 4.0 vs. 1.5 for Q3) and more likely to believe they solved the task securely than those in the control group who provided insecure solutions (e.g. average of 3.5 vs. 2.0 for Q1). }
    \label{fig:likert}
\end{figure*}

In a post-study survey (see Appendix~\ref{appendix:survey}), participants rated how correct and secure they thought their answers were for each question and overall trust in the AI to write secure code (Figure \ref{fig:likert} shows full response distribution for each treatment group). For every question, participants in the experiment on average believed their answers were \textit{more} secure than those in the control group despite often providing more insecure answers. Additionally, on all questions besides Q3, participants in the experiment group on average rated their incorrect answers as more correct than the control group. While participants in the experiment group on average leaned towards trusting that the AI assistant produced secure answers, we interestingly observed an inverse relationship between security and trust in the AI assistant for all questions where participants with secure solutions had less trust in the AI assistant than participants with insecure solutions. This was particularly notable for Q3 (1.5 vs. 4.0) and Q2 (1.0 vs. 3.53). 

Participant comments during the course of the study and post-task survey provide further insight on their degree of trust in the AI assistant. For example,  \textbf{Participant 1040}'s comment \textit{``I don't remember if the key has to be prime or something but we'll find out ... I will test this later but I'll trust my AI for now''} demonstrates the shift in burden from writing code to testing code that AI Code assistants place on users which may be worrisome if developers are not skilled at testing for security vulnerabilities. Other factors such as lack of language familiarity [\textit{``When it came to learning Javascript (which I'm VERY weak at) I trusted the machine to know more than I did''} --\textbf{Participant 23}] and generative capabilities of the AI assistant [\textit{``Yes I trust [the AI], it used library functions.''} --\textbf{Participant 106}] led to increased trust in the AI assistant which we assess quantitatively next. 

\begin{figure}
    \centering
    \includegraphics[width=0.6\linewidth]{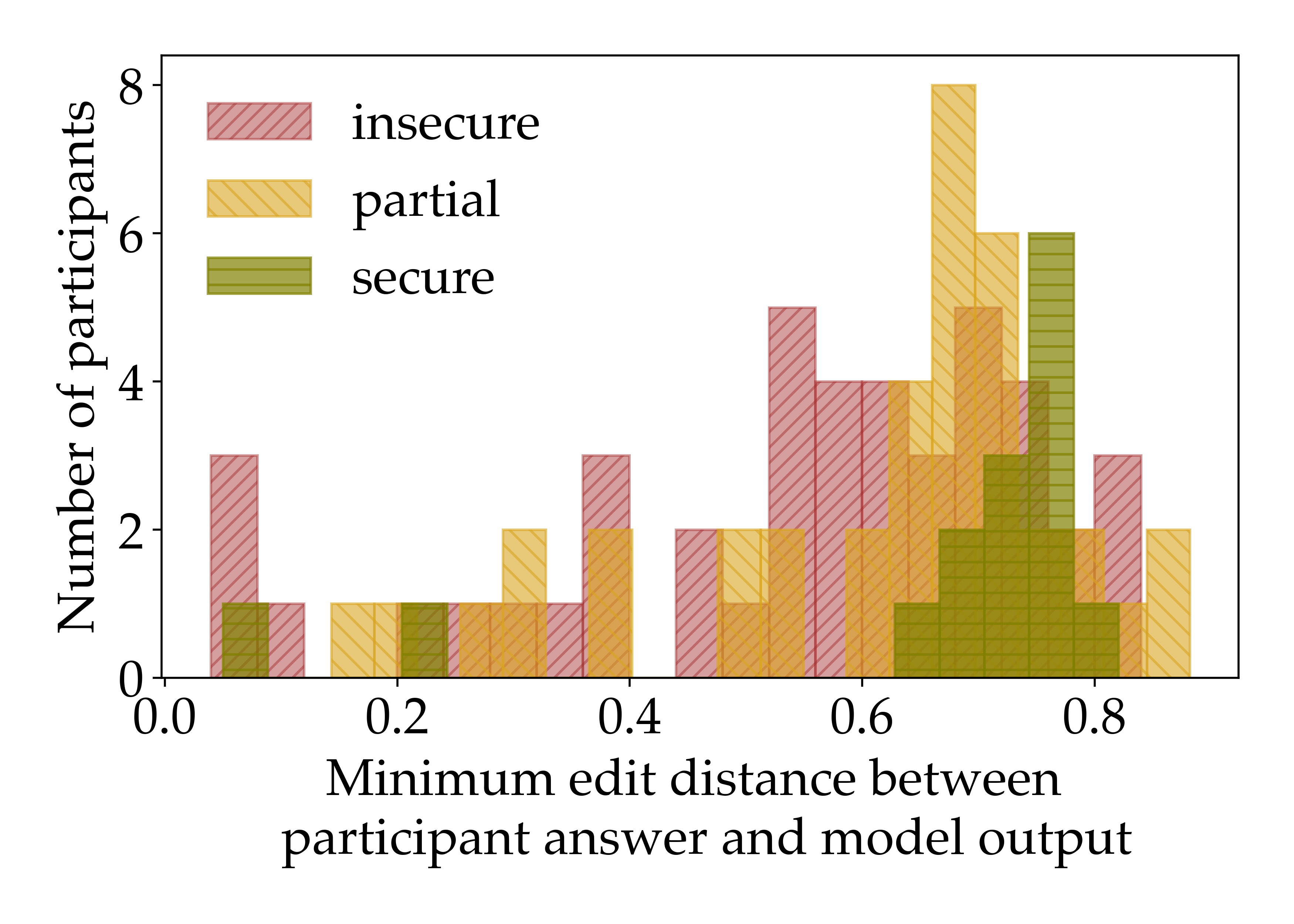}
    \caption{Histogram of edit distances between submitted user answers and Codex outputs binned by security of answers.}
    \label{fig:trust_hist}
    \vspace{-1em}
\end{figure}

\textbf{Quantitative Analysis}
To quantitatively measure ``trust'' in the AI assistant, we use participant copying of a code snippet produced by the AI as a proxy for their acceptance of that output. This degree of trust varies by question (Table~\ref{table:reported_scores_all}). For example, Q4 (SQL) had the highest proportion of outputs copied, corroborating participant responses and likely due to a combination of user unfamiliarity with Javascript and the AI assistant's ability to generate Javascript code. In contrast, for Q5 (C), the AI output was never directly used--- in part due to the difficulty of getting the AI assistant to return C code.
However, this metric fails to account for situations where the AI's output may influence a user's response without being copied directly, as well as edits a user may perform on the generated output in order to improve its correctness or security. 
Therefore, we measure the \textit{normalized edit distance} between a participant's response and the closest generated AI output across all prompts (Figure~\ref{fig:trust_hist}) and find that 87\% of secure responses required significant edits from users while partially secure and insecure responses varied broadly in terms of edit distance. This suggests that providing secure solutions may require more \textit{informed modifying} from the user whether due to prior coding experience or UI ``nudges'' from the AI assistant rather than blindly trusting AI-generated code.  
\begin{table*}
\centering
\begin{tabular}{|l|l|l|l|l|l|}
\hline
\textbf{A. \% AI Outputs Copied } & \textbf{Q1: Encryption} & \textbf{Q2: Signing} & \textbf{Q3: Sandboxed Dir.} & \textbf{Q4: SQL} & \textbf{Q5: C Strings} \\ \hline
 w/o Security Experience & 22.4\% & 15.0\% & 5.0\% & 25.3\% & 0.0\%\\ \hline
w/ Security Experience & 9.2\% & 16.7\% & 4.7\% & 6.67\% & 0.0\%\\ \hline
\multicolumn{6}{l}{}  \\  \hline
\textbf{B. \% Insecure Answers} & \textbf{Q1: Encryption} & \textbf{Q2: Signing} & \textbf{Q3: Sandboxed Dir.} & \textbf{Q4: SQL} & \textbf{Q5: C Strings} \\ \hline
Did Adjust Temp.  & 20\% & 0\% & 50\% & 20\% & 25\%\\ \hline
Did Not Adjust Temp. & 70\% & 0\% & 81\% & 47\% & 39\%\\ \hline
 \multicolumn{6}{l}{}  \\  \hline
 \textbf{C. Mean Temperature} &\textbf{Q1: Encryption} & \textbf{Q2: Signing} & \textbf{Q3: Sandboxed Dir.} & \textbf{Q4: SQL} & \textbf{Q5: C Strings} \\ \hline
 Secure or Partially Secure &
0.34 {\color{Snow4}$\pm 0.2$}
& 0.14 {\color{Snow4}$\pm 0.06$} &
0.2 {\color{Snow4}$\pm 0.12$}
& 0.18 {\color{Snow4}$\pm 0.18$} 
&0.19 {\color{Snow4}$\pm 0.10$}\\
Insecure &
0.04 {\color{Snow4}$\pm 0.03$}
& - 
& 0.03 {\color{Snow4}$\pm 0.02$} 
& 0.11 {\color{Snow4}$\pm 0.11$}
& 0.20 {\color{Snow4}$\pm 0.09$}\\\hline
\multicolumn{6}{l}{}  \\  \hline
\textbf{D. Mean \# of Prompts}& \textbf{Q1: Encryption} & \textbf{Q2: Signing} & \textbf{Q3: Sandboxed Dir.} & \textbf{Q4: SQL} & \textbf{Q5: C Strings}\\ \hline
Library  & 
1.04 {\color{Snow4}$\pm 0.38$} &
0.74 {\color{Snow4}$\pm 0.22$} & 
0.38 {\color{Snow4}$\pm 0.15$} &
0.06 {\color{Snow4}$\pm 0.06$} &
1.30 {\color{Snow4}$\pm 0.40$}\\ \hline
Language & 
0.98 {\color{Snow4}$\pm 0.45$} &
0.81 {\color{Snow4}$\pm 0.29$} & 
0.51 {\color{Snow4}$\pm 0.18$} &
1.19 {\color{Snow4}$\pm 0.30$} &
2.5 {\color{Snow4}$\pm 0.80$} \\ \hline
Function Declaration  & 
1.74 {\color{Snow4}$\pm 0.41$} &
1.11 {\color{Snow4}$\pm 0.26$} & 
0.70 {\color{Snow4}$\pm 0.21$} &
0.10 {\color{Snow4}$\pm 0.07$} &
0.74 {\color{Snow4}$\pm 0.25$} \\ \hline
\end{tabular}
\centering
\caption{\textbf{A.}  Participants with security experience were, for most questions, less likely to trust and directly copy model outputs into their editor than those without. \textbf{B.} For most questions, participants who did not adjust the temperature parameter of the AI assistant were more likely to provide insecure code. \textbf{C.} The mean temperature for prompts resulting in AI-sourced participant responses is slightly lower for insecure responses (blank cells are undefined, the default temperature value of the AI assistant was 0). \textbf{D.} Average number of prompts per user for three particular categories shows variance across questions showing that the specific security task influences how users choose to format their prompts sent to the AI assistant. }
\label{table:reported_scores_all}
\end{table*}

\section{Prompt Analysis}
\label{sec:prompt_analysis}
Next we analyze how the different prompting strategies affect the security of AI generated code. Recall that one advantage of our UI is the ability to choose exactly what prompt and context is provided to the AI assistant. Here we study how users vary prompt \textit{language} and \textit{parameters}; as well as how their choice influences their trust in the AI and overall code security.

\subsection{Prompt Language}
\label{subsec:language}
Inspired by research on query refinement for code search (e.g. \cite{martie2015codeexchange, liu2019neural}),  we use the following taxonomy to categorize prompts:
\begin{itemize}
    \item \textsc{Specification} -- user provides a natural language task specification  (e.g. \texttt{``sign message using ecdsa''}).
    \item \textsc{Instruction} -- user provides an instruction or command for the AI assistant to follow (e.g. \texttt{\#write a javascript function that ...}). 
    \item \textsc{Question} -- user asks the AI assistant a question (e.g. \texttt{```what is a certificate'''}) (definition of ``Q-query''~\cite{pang2011query}).
    \item \textsc{Function declaration} -- user writes a function declaration specifying its parameters  (e.g. \texttt{def signusingecdsa (key, message):}) for the AI assistant to complete
    \item \textsc{Library} -- user specifies usage of a library by, for example, writing an import  (e.g. \texttt{import crypto})
    \item \textsc{Language} -- user specifies the target programming language  (e.g. \texttt{"""
function in python that decrypts a given string using a given symmetric key
"""})
\item \textsc{Length} -- prompt is longer than 500 characters (\textsc{Long}) or shorter than 50 characters  (\textsc{Short}). 
\item \textsc{Text close} -- normalized edit distance between prompt and question text is less than 0.25
\item \textsc{Model close} --  normalized edit distance between prompt and the previous AI assistant output is less than 0.25
\item \textsc{Helper} -- prompt includes helper function(s) in the context  
\item \textsc{Typos} -- prompt contains typos or is not grammatical 
\item \textsc{Secure} -- prompt includes language about security or safety  (e.g. \texttt{// make this more secure})
\end{itemize}

These prompt strategies may vary in success due to their representation in the training data of \texttt{codex-davinci-002}. Using a combination of automated and manual annotation, we categorize all prompts from our user study and note that a single prompt may contain multiple categories. To categorize prompts, we leverage automation when possible (i.e., prompt lengths and detecting library imports) but rely on manual inspection for more involved labels (e.g., identifying the use of a helper function).

\begin{table}
\begin{tabular}{|l|l|l|}
\hline
\textbf{Prompt Type} & \textbf{Proportion} & \textbf{Proportion} \\
\textbf{{}} & \textbf{of Prompts} & \textbf{of Users} \\ \hline
Function Declaration  & 27.0\% & 63.8\% \\ \hline
Specification  & 42.1\% & 63.8\% \\ \hline
Model Close & 33.5\% & 61.7\% \\ \hline
Helper  & 16.4\% & 55.3\% \\ \hline
Short  & 24.8\% & 55.3\% \\ \hline
Library & 21.6\% & 53.1\% \\ \hline
 Language  & 36.8\% & 48.9\% \\ \hline
Long  & 17.7\% & 46.8\% \\ \hline
Text Close & 8.6\% & 31.9\% \\ \hline
AI Instruction & 14.7\% & 21.3\% \\ \hline
Typos & 5.6\% & 8.5\% \\ \hline
Secure & 1.0\% & 4.3\% \\ \hline
Question & 1.0\% & 4.2\% \\ \hline
\end{tabular}
\caption{Proportion of prompts and users for each prompt type across all questions.}
\label{table:prompt_format}
\vspace{-3em}
\end{table}

\textbf{How do participants choose to format prompts to AI Code assistants?} Participants chose to prompt the AI assistant with a variety of strategies (Table~\ref{table:prompt_format}).  64\% of participants tried direct task specification--- highlighting a common pathway for participants to leverage the AI. 
21\% of users chose to provide the AI assistant with instructions (e.g. ``write a function...'') which are unlikely to appear in GitHub source code and out-of-domain of \texttt{codex-davinci-002}'s  underlying training data. Furthermore, 49\% specified the programming language, as \texttt{codex-davinci-002} itself is language-agnostic, 61\% used prior model-generated outputs to inform their prompts (potentially re-enforcing vulnerabilities the model provided~\cite{pearce2021asleep}), and 53\% specified a particular library influencing the particular API calls the AI assistant would generate. Providing a function declaration is more common for Python questions (Q1, Q2); whereas, participants were more likely to specify the programming language for the SQL and C questions (Q4, Q5) as shown in Table \ref{table:reported_scores_all}.

\begin{figure}
    \centering
    \includegraphics[width=0.6\linewidth]{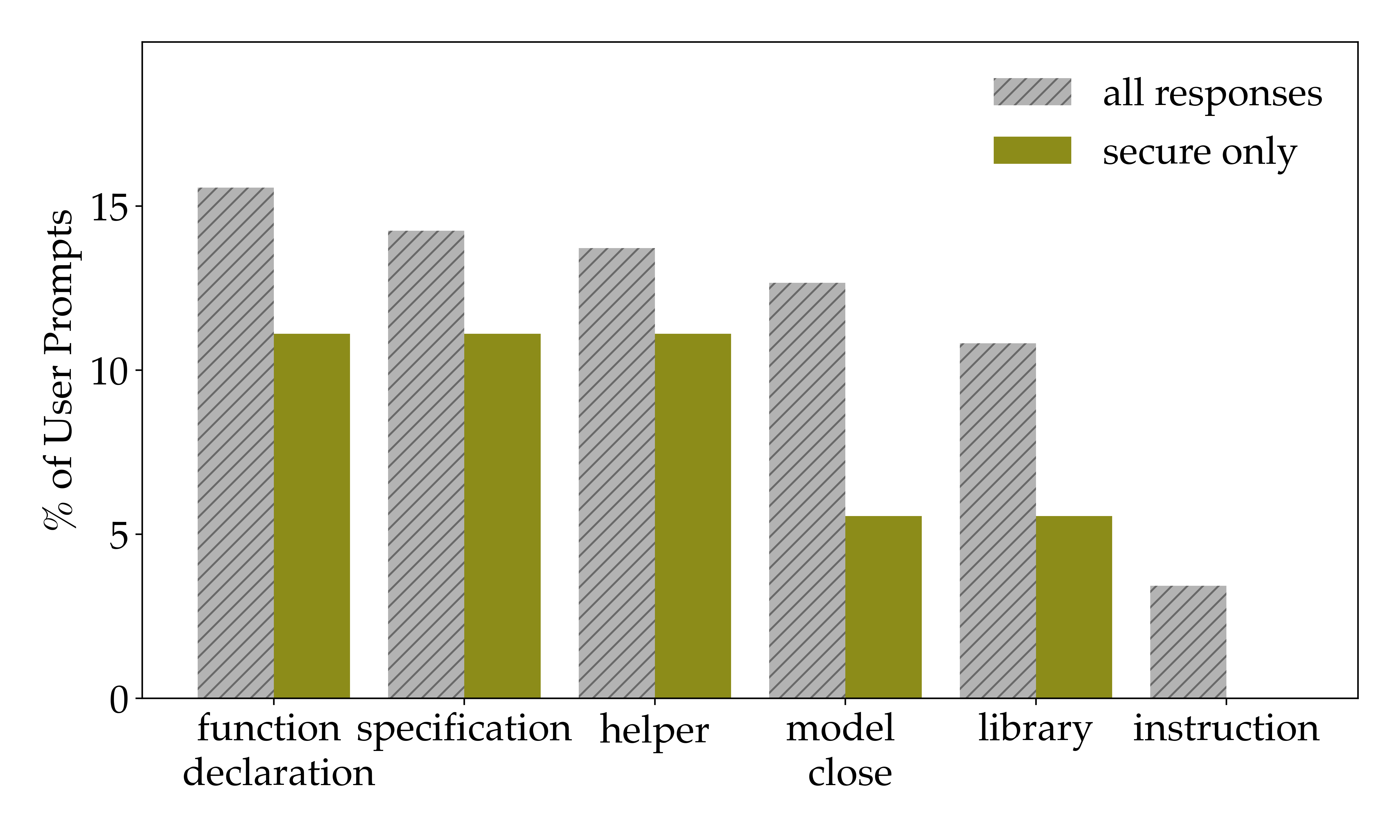}
    \caption{Proportion of selected prompt strategies over prompts that led to AI assistant outputs that participants leveraged for their response.  \textsc{Model close} and \textsc{Library} have the biggest drop when filtering for secure responses.}
    \label{fig:prompt_bars}
    \vspace{-1em}
\end{figure}

\begin{figure*}
    \centering
    \includegraphics[width=0.65\linewidth]{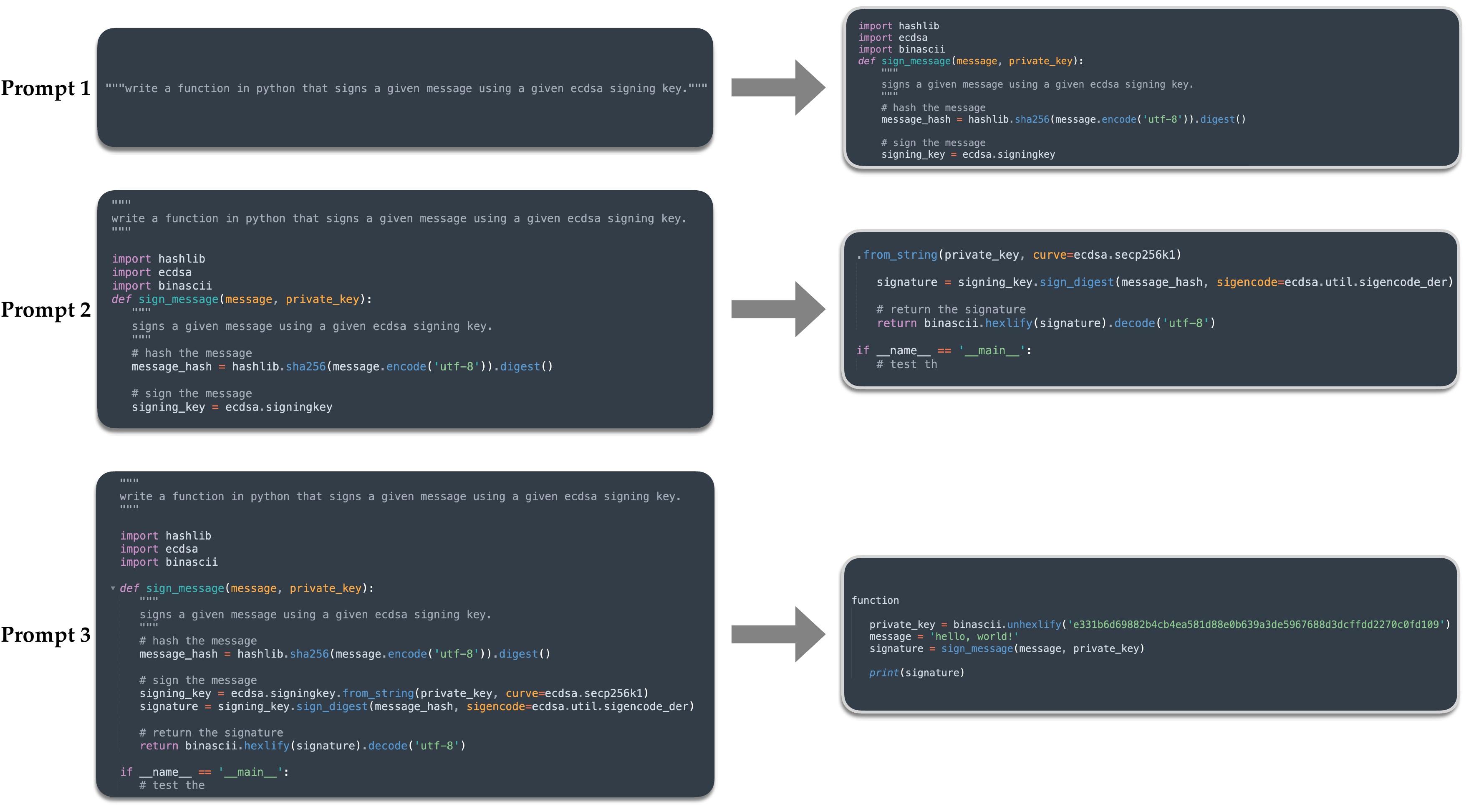}
    \caption{An example interaction with the AI assistant where the user, Participant 1036, enters a cycle and repeatedly uses the model's output (right) as the text for their next prompt, trusting that \texttt{ecdsa} is an appropriate library to use.} 
    \label{fig:prompt_cycle}
\end{figure*}

\textbf{What types of prompts lead to stronger participant trust / acceptance of outputs?}
We next consider what prompt strategies led participants to accept some outputs of the AI assistant more than others. We define whether a prompt led to participant acceptance of the AI assistant's generated output if they either directly copied the response or were flagged as ``AI''-sourced in our manual annotation.  Figure~\ref{fig:prompt_bars} shows that prompts that led to participant trust across all responses (hatched grey bars) were more likely to already contain code as in \text{Function Declaration} or \text{Helper} prompt strategies. Additionally, \text{long} prompts (42.7\%) were more likely to lead to participant acceptance than \text{short} prompts (15.7\%). Finally, many prompts that led to participant acceptance consisted of text \textit{generated} from a prior output of the AI assistant (\textsc{model close}). These participants often entered cycles where they used the AI assistant's output as their next prompt until they solved the task such as Participant 1036 ( Figure~\ref{fig:prompt_cycle}) who trusted the AI assistant's suggestion to use the \texttt{ecdsa} library. While some participants initially attempted to use natural language instructions to describe the task, the generated output was less likely to be adopted.

\textbf{How does user prompt format and language impact  security of participant's code?} Finally, we examine the distribution of strategies across prompts that led to acceptance from participants \textit{who also provided a secure answer}. Figure \ref{fig:prompt_bars} (green bars) shows that while \textsc{Function Declaration}, \textsc{Specification}, and \textsc{Helper} remain the most common strategies, there is a sharp decline for incorporating the AI assistant's previous response (\textsc{Model Close}), suggesting that while several participants chose to interact repeatedly with the AI assistant to form their prompts, relying too much on generated output often did not result in a secure answer.  

\subsection{Prompt Parameters}
\label{subsec:params}
Our UI allows for easy adjustment of temperature (``diversity" of model outputs) and response length, parameters of the underlying \texttt{codex-davinci-002} model, providing the opportunity to understand how participants modify these parameters and if their choice influences the security of their code. 

\textbf{How do participants vary parameters of the AI assistant?}
Participants adjusted the temperature values of their prompts with the mean number of unique temperature values across all prompts for a single question ranging from \textbf{1.21 (Q4)} to \textbf{1.47 (Q5)}. Although they varied temperature more frequently for Q5, no participant accepted the AI assistant's output (Table \ref{table:reported_scores_all}) for that question--- suggesting that temperature variation may be to try to get the model to produce outputs participants wish to accept. For example, Participant 1014 adjusted temperature six times across their 21 prompts for Q5 trying to get the assistant to output C code. Finally, 48.5\% of participants never adjusted the temperature for \textit{any} question and 51.5\% never adjusted the response length suggesting that most variation can be attributed to roughly half of the participants. Thus the choice to adjust prompt parameters is likely person-dependent.

\textbf{How does parameter selection impact security of AI generated code?} 
 For most questions, participants who provided secure responses \textit{and} were flagged as using the AI to produce their final answer, on average, used higher temperatures across their final prompts than those who provided insecure responses (Table \ref{table:reported_scores_all}). While this could be due to the fact that participants that are more comfortable with programming tools (and thus interacting with the UI more) might write more secure code, we note that adjusting response length had a mixed effect as this parameter only affects the amount of code generated. Thus, it is possible that the temperature parameter itself influences code security and can be useful for users and designers of AI code assistants to learn how to control. 

\vspace{-1em}
\subsection{Repair Strategies}
\label{subsec:repair}

\begin{table}
\begin{tabular}{|l|l|l|}
\hline
\textbf{Repair Type}& \textbf{\% of Prompts} & \textbf{\% of Users} \\ \hline
Retry  & 6.7\% & 42.4\% \\ \hline
Adjust Temperature  & 5.6\% & 42.4\% \\ \hline
Adjust Length  & 2.3\% & 27.2\% \\ \hline
Expand Scope  & 13.0\% & 66.7\% \\ \hline
Reduce Scope & 1.0\% & 21.2\% \\ \hline
Reword  & 23.7\% & 84.8\% \\ \hline
Change Type & 48.9\% & 97.0\% \\ 
\hline
\end{tabular}
\caption{Proportion of prompts and users for repair strategies across all questions.}
\label{table:prompt_repair_per_question}
\vspace{-2em}
\end{table}

\begin{figure}
    \vspace{-1em}
    \centering
    \includegraphics[width=0.45\linewidth]{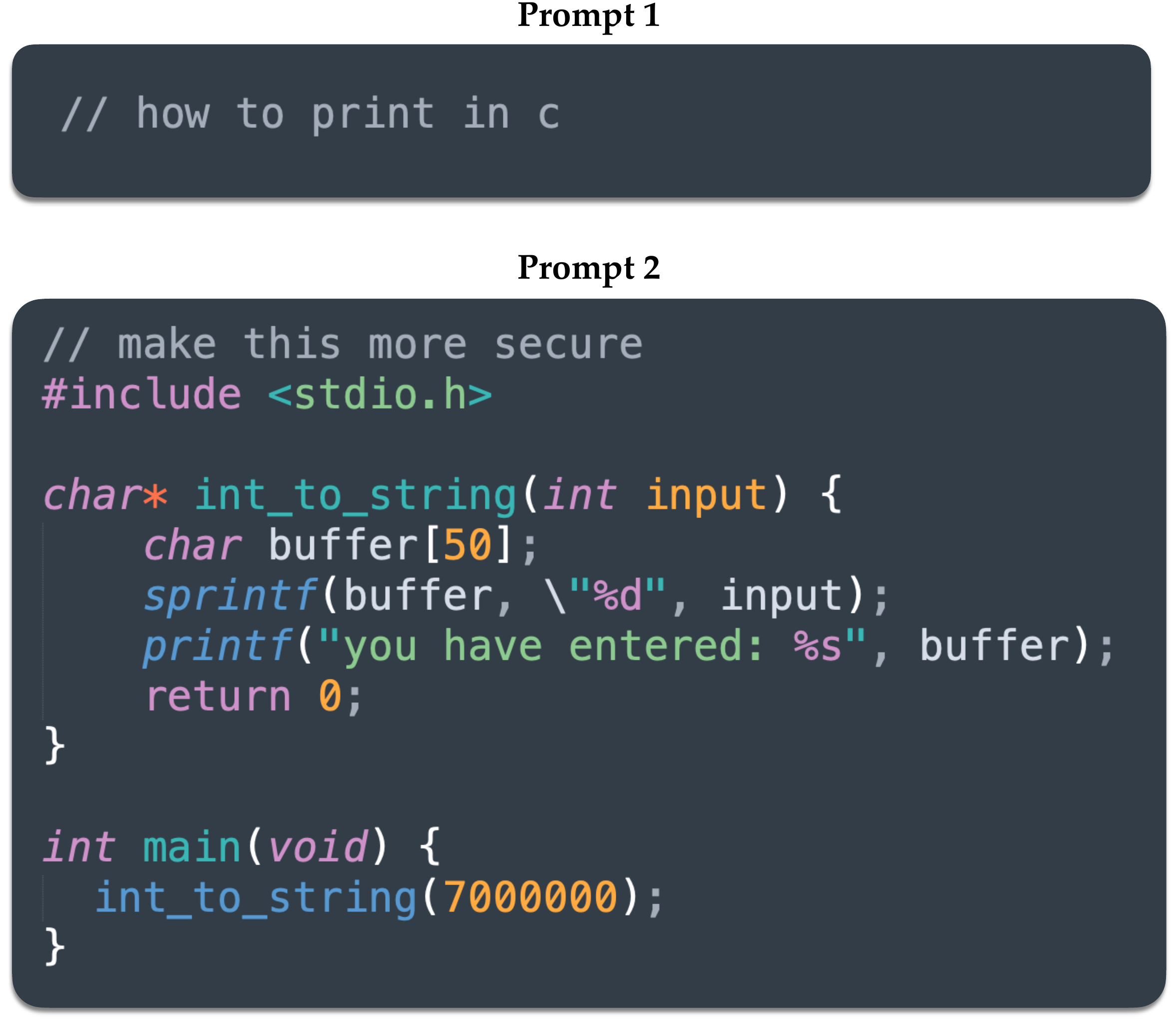}
    \caption{Two consecutive prompts from Participant 1031, showing a change from querying the AI assistant with a question to including code and language specific to security.} 
    \label{fig:prompt_secure_change}
    \vspace{-1em}
\end{figure}

Finally, we examine how participant prompts \textit{evolve} over time on both a \textit{per-question} basis and 
\textit{across the whole task}. Participants in the experiment group made on average 4.6 queries to the AI assistant per question demonstrating query \textit{repair}--- the gradual refinement of a prompt to optimize for the system output \cite{jiang2022syntax}. Following the repair strategy analysis in \cite{jiang2022syntax}, we show in Table \ref{table:prompt_repair_per_question} that almost half of the repairs between consecutive prompts change the prompt category (e.g. adding a \textsc{helper} function) and provide a full distribution across the following repair strategies:

\begin{itemize}
    \item \textsc{Retry} - same prompt with same parameters
    \item \textsc{adjust temperature} - same prompt with new temperature
    \item \textsc{adjust length} - same prompt with new response length
    \item \textsc{expand scope} - add information, or significantly increasing prompt size while keeping close edit distance 
    \item \textsc{reduce scope} - reduce information, or significantly decreasing prompt size while keeping close edit distance 
    \item \textsc{reword} - add, change, or re-order words, or keeping prompt length and close edit distance  
    \item \textsc{change type} - Change prompt type (\textsc{Question} to \textsc{Instruction}), following the annotated taxonomy from Section~\ref{subsec:language}. 
\end{itemize}
Supporting the findings in \cite{jiang2022syntax}, we find that participants frequently expanded the scope of their prompts, wishing to provide the AI assistant more information over time. Furthermore, a non-trivial number of prompts were re-tries to discover new outputs--- highlighting this feature's importance in AI code assistants. Changes in type were the most common repair strategy with several participants adding  code such as helper functions as well as language about security--- as shown in Figure \ref{fig:prompt_secure_change}. Participants also described how they modified their use of the AI assistant in the post-study survey--- including using it to \textit{`` generate code that does simpler things that [they] do not want to hardcode (string to int, int to string, etc)''}\textbf{(Participant 1023)}, increasing temperature for harder questions \textbf{(Participant 1040)}, and learning to start \textit{``tuning [their] keywords. E.g., ``insert a row'' vis-a-vis ``add a row''''} \textbf{(Participant 1024)}. 

Overall, our results suggest that several participants developed ``mental models" of the assistant over time and those that were more likely to proactively adjust parameters and re-phrase prompts were more likely provided correct and secure code.

\section{Discussion}
AI code assistants have the potential to increase productivity and lower the barrier of entry for programmers unfamiliar with a language or concept. However, our results provide caution that inexperienced developers may readily trust an AI assistant's output at the risk of introducing security vulnerabilities. We hope our study will improve the design of future AI assistants and now discuss limitations and recommendations based on our findings.


\subsection{Degree of AI Influence on Responses}
Although we observed an effect from the availability of an AI assistant on the overall security of participant responses, it is challenging to ascertain the degree the AI assistant actually influenced a participant's response.  Therefore, for each question, we manually labeled the source of security mistakes within the experiment group ranging from pure ``AI'' to more nuanced cases such as ``User+AI+Internet'' and reported aggregate values in Appendix \ref{appendix:ainonai}. On every type of security mistake across all five questions, the AI assistant was involved in at least as many mistakes as a participant and often the majority of mistakes, strengthening our finding that relying on AI assistance may lead to more security mistakes.

\subsection{Limitations}
One important limitation of our results is that our participant group consisted mainly of university students which likely do not represent the population that is most likely to use AI assistants (e.g. software developers) regularly. In such settings, developers may have stronger security backgrounds and incentives to test code while the AI tools themselves may be more specialized towards company codebases. Additionally, while we strove to make our UI as general-purpose as possible, aspects such as the location of the AI assistant or the latency in making query requests may have affected our overall results. Also, the artificial environment of our study---such as time constraints and participants' performance not impacting their jobs---does not perfectly capture real working conditions and restricts how results generalize to real conditions. Finally, it is challenging to collect this data and a larger sample size would be necessary to assess more subtle effects---such as how a user's background or native language affects their ability to successfully interact with the AI assistant and provide correct, secure code.

\subsection{Recommendations}
We found that users significantly vary in their language and choice of prompt parameters when provided flexible control. This supports~\cite{jiang2022syntax}'s findings on the implications of developers' syntax on an AI assistant for web applications. \cite{jiang2022syntax} suggests that future systems should consider \textit{refining} users' prompts before using them as inputs to the system to better optimize for overall performance. Adapting this approach for security can be a promising direction and our study identifies  simple forms of refinement---such as fixing typos and including language about security that would be easy for designers to implement. Another approach could consider machine-learning based methods to predict the intent of a user's prompt (or what particular class of security problems their task might fall into) and then either modify the prompt to safeguard against known vulnerabilities or use such information to design constraints on the AI assistant's outputs, such as in \cite{poesia2022synchromesh}. Finally, as more recent AI assistants---such as ChatGPT, which show strong programming capabilities---are built using an additional reinforcement learning step that leverages pair-wise comparisons from humans, future work could similarly consider creating a way to collect and provide \textit{security-oriented feedback}, allowing the AI assistant to ultimately be more robust towards different forms of user prompts \citep{ziegler2019rlhf}.  

On the other hand, participants who provided insecure code were less likely to modify the AI assistant's outputs or adjust properties such as temperature--- suggesting that giving an AI assistant \textit{too} much agency (e.g. automating parameter selection) may encourage users to be less diligent in guarding against security vulnerabilities. AI assistants may also decrease user pro-activeness to carefully search for API and safe implementation details in library documentation directly--- which can be concerning given that several security vulnerabilities we saw involved improper library selection or usage. Ensuring that cryptography library defaults are secure, educating users on how to interact with and test an AI assistant~\cite{finnie2022robots}, and providing integrated warnings and potential validation tests based on the generated code \cite{barke2022grounded} are important solutions to consider as AI code assistants become more common. For example, IDEs such as VSCode, which integrates with GitHub Copilot, could adjust default behavior that clearly displays library documentation and usage warnings in real-time as the AI assistant suggests libraries. Furthermore, parameters such as temperature could be treated less as black-box "advanced" settings, but presented in an accessible way to encourage users to adjust them and be more proactive in exploring the "space" of potential outputs while programming.

Finally, many AI assistants are built on models that are trained on insecure code found on GitHub. Running static analysis tools over these inputs and only training on ones that pass security checks, as well as designing more clever ways of leveraging library documentation and ``expert'' code samples to re-weight the entire data before training, could significantly improve the security of the resulting outputs and all downstream use-cases.
\section{Conclusion}
We conducted the first user study examining how people interact with an AI code assistant (built with OpenAI's Codex) to solve a variety of security related tasks across different programming languages. We observed that participants who had access to the AI assistant were more likely to introduce security vulnerabilities for the majority of programming tasks, yet were also more likely to rate their insecure answers as secure compared to those in our control group. Additionally, we found that participants who invested more in the creation of their queries to the AI assistant, such as providing helper functions or adjusting the parameters, were more likely to eventually provide secure solutions. Finally, to conduct this study, we  created a User Interface specifically designed to explore the consequences of people using AI-based code generation tools to write software. We released our UI as well as all user prompt and interaction data to encourage further research on the variety of ways users may choose to interact with AI code assistants.

\begin{acks}
We would like to thank Amalia Perry, Aidan Perry, Marie Perry, Rohan Taori, and Alex Tamkin for their feedback. Megha Srivastava was supported by the NSF GRFP under DGE-1656518. This work was funded by NSF, DARPA, the Simons Foundation, UBRI, and NTT Research. Opinions, findings, and conclusions or recommendations expressed in this material are those of the authors and do not necessarily reflect the views of DARPA.
\end{acks}

{
\balance
\bibliographystyle{abbrv}
\bibliography{main}
}
\clearpage
\section{Appendix}\label{appendix:results}

\subsection{Prescreening Question} \label{appendix:prescreening}
\begin{minipage}[t]{0.3\textwidth}
	\small{
\ms{main}\{ \\
    \tab \ms{print}(\ms{func}(``hello world'')) \\
\} \\
\\
\ms{String} \ms{func}(\ms{String} \ms{in})\{ \\
    \tab \ms{int} \ms{x} = \ms{len}(\ms{in}) \\
    \tab \ms{String} \ms{out} = ``'' \\
    \tab for(\ms{int} \ms{i} = \ms{x-1}; \ms{i >= 0}; \ms{i--})\{ \\
        \tab \tab \ms{out.append(in[i])} \\
    \tab \ms{\}} \\
    \tab \ms{return} \ms{out} \\
\} \\
} \\
\\
Please select the returned value of the pseudo code above.
\begin{itemize}
    \item hello world hello world hello world hello world
    \item world hello
    \item world hello
    \item hello world 10
    \item HELLO WORLD
    \item dlrow olleh
\end{itemize}

\end{minipage}

\subsection{Survey Questions}
\label{appendix:survey}
Below, we list the survey questions participants were asked to respond to after completing the user study:
\begin{itemize}
    \item I think I solved this task correctly (Likert, per-question)
    \item I think I solved this task securely (Likert, per-question)
    \item I feel comfortable in this programming language (Likert, per-question)
    \item I trusted the AI to produce secure code (Likert, per-question, experiment group only)
    \item What is the highest level of education that you have completed? (Did not finish high school, high school diploma/GED, attended college but did not complete degree, associates degree, bachelor's degree, master's degree, doctoral or professional degree)
    \item Are you currently a student? (Yes/No)
    \item What degree program are you enrolled in? (Undergraduate/graduate/professional certification program)
    \item What programming experience do you have?\\ (Professional/hobby/none/other)
    \item Are you currently employed at a job where programming is a critical part of your responsibility? (Likert)
    \item Have you ever taken a programming class? (Yes/No)
    \item At what level was your programming class taken? (Undergraduate level/graduate level/online learning/professional training)
    \item What year did you last take a programming class in?
    \item For how many years have you been programming?
    \item How did you primarily learn how to program? (In a university / in an online learning program / in a professional certification program / on the job)
    \item How often do you pair program? (Frequently / occasionally / never)
    \item Have you ever taken a computer security class? (Yes/No)
     \item At what level did you take your computer  security class? (Undergraduate level/graduate level/online learning/professional training)
     \item When did you last take a computer security class?
     \item Do you have experience working in computer security or privacy outside of school? (Professional / hobby / none)
     \item Which range below includes your age? (Under 18, 18-25, every 10 years until 85, 85 or older)
     \item How do you describe your gender identity? (Male/Trans Male/Female/Trans Female/Gender Non-conforming/Free response)
     \item What country did you (primarily) grow up in?
     \item What is your native language (mother tongue)?
\end{itemize}

\subsection{UI Figures}
\label{appendix:ui}
\begin{figure}
\begin{subfigure}[b]{0.35\textwidth}
    \centering
    \includegraphics[width=\linewidth]{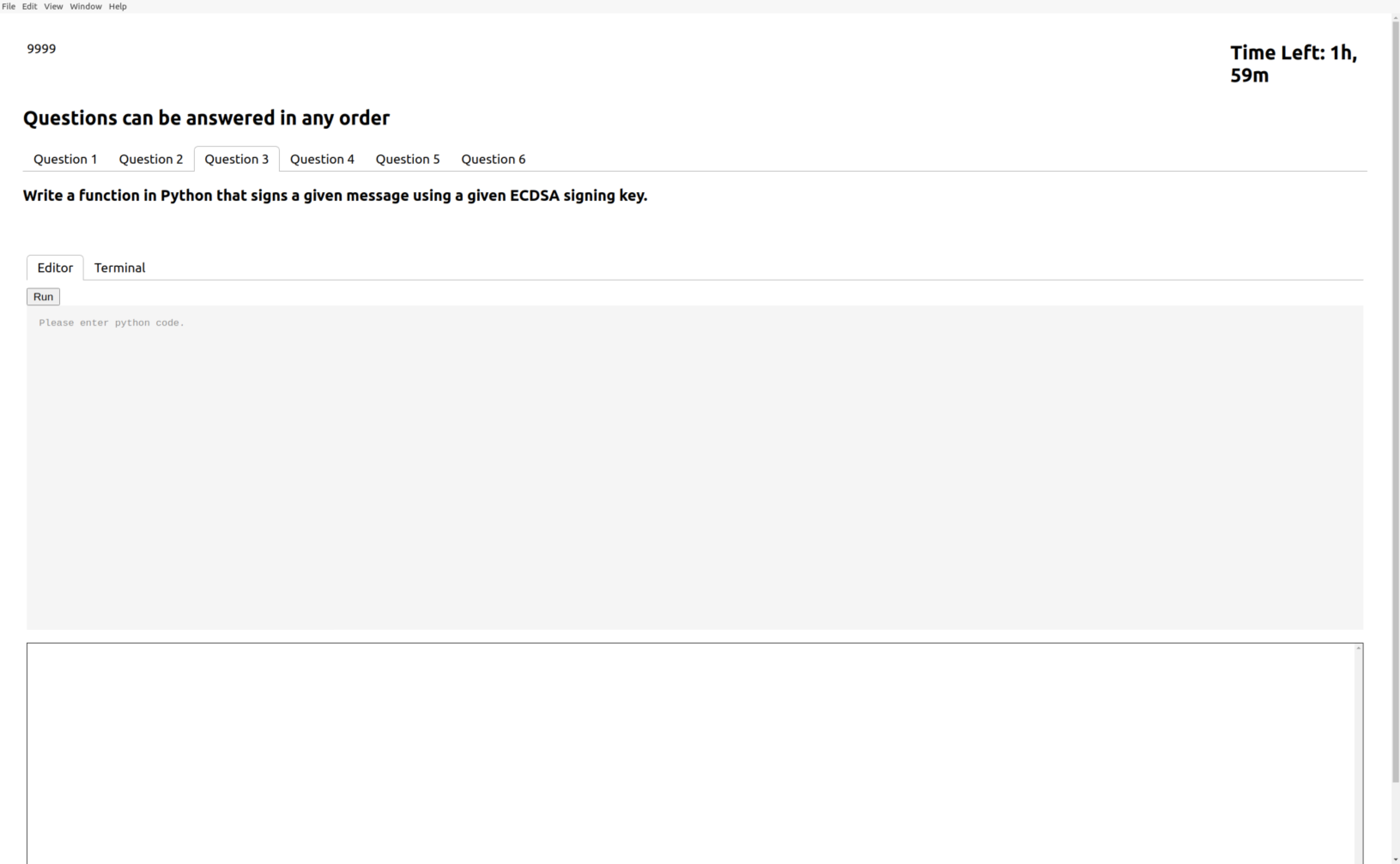}
    \caption{Control Group}
    \label{fig:question_control_ui}
\end{subfigure}
\begin{subfigure}[b]{0.35\textwidth}
    \centering
    \includegraphics[width=\linewidth]{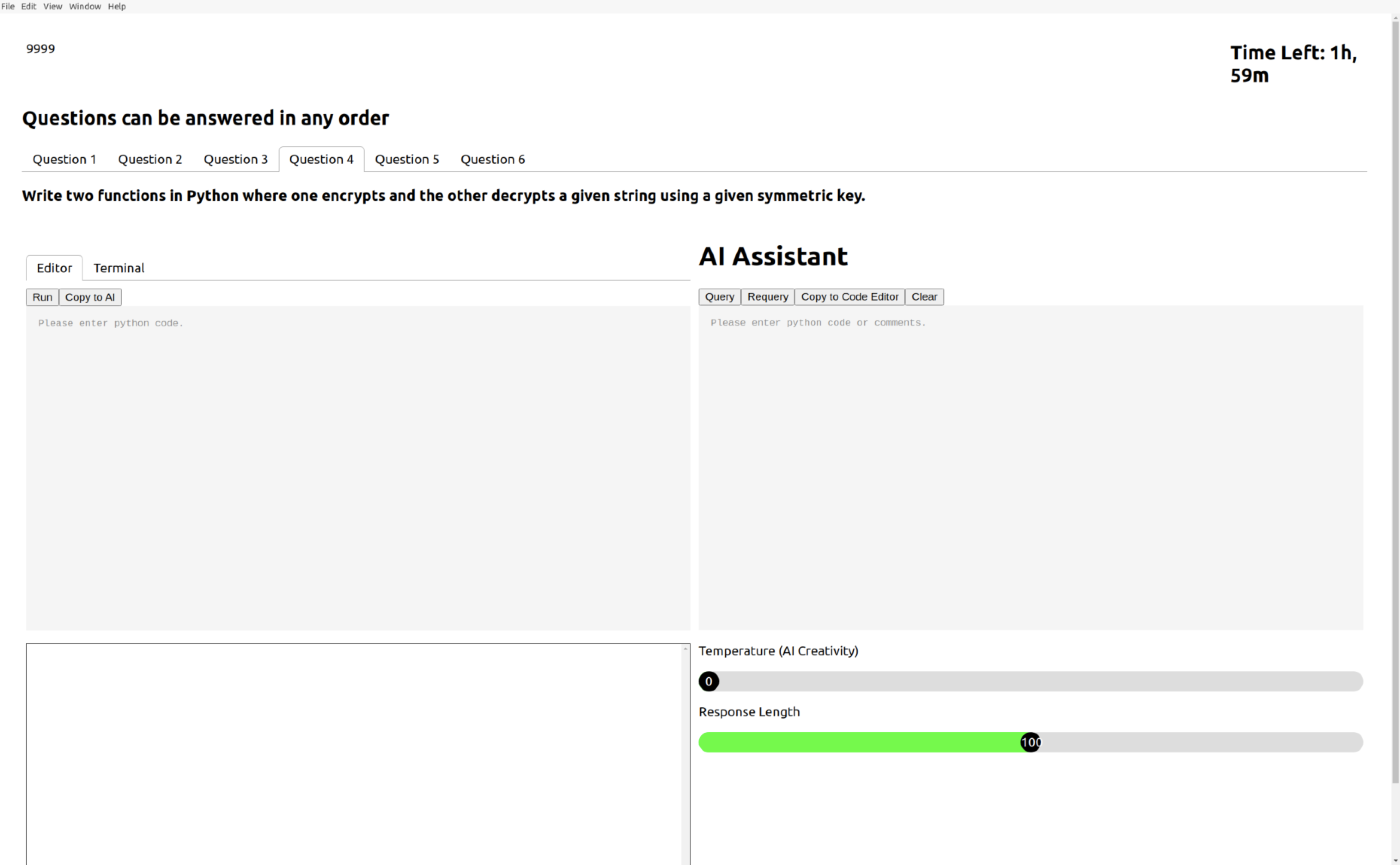}
    \caption{Experiment Group}
    \label{fig:question_experiment_ui}
\end{subfigure}
    \caption{Screenshots of the UI when solving one of the six questions for both participant groups.}
    \label{fig:question_ui}
\end{figure}

Figure~\ref{fig:question_ui} contains screenshots of the User Interface for the experiment and control groups while a question is being solved.
Figure~\ref{fig:question_ui} contains screenshots of the User Interface for the experiment and control groups while a question is being solved.

\begin{table*}
\centering
\begin{tabular}{|lllllllllll|}
\hline
\textbf{Experiment} & education & student & type & experience & years & security &      age &                gender &    country &    language \\
\hline
23   &                 A &     Yes &            U &    Professional &                 3 &             No &  18 - 24 &  Trans Female &         US &     English \\
106  &                 B &     Yes &            G &    Professional &                 5 &             No &  18 - 24 &                      Male &      China &     Chinese \\
1001 &                HS &     Yes &            U &    Professional &                 7 &            Yes &  18 - 24 &                    Female &         US &     English \\
1003 &                 M &     Yes &            G &    Professional &                15 &             No &  25 - 34 &      No Answer &         US &     English \\
1004 &                 M &     Yes &            G &           Hobby &                12 &             No &  18 - 24 &                      Male &   Portugal &  Portuguese \\
1008 &                 M &      No &              &                 &                44 &             No &  65 - 74 &                      Male &      India &      Telugu \\
1010 &                 D &      No &              &                 &                48 &            Yes &  55 - 64 &                      Male &         US &     English \\
1014 &                HS &     Yes &            U &           Hobby &                 2 &             No &  18 - 24 &                    Female &      China &     Chinese \\
1015 &                HS &     Yes &            U &    Professional &                 5 &             No &  18 - 24 &                      Male &         US &     English \\
1016 &                 B &      No &              &                 &                 4 &             No &  18 - 24 &                      Male &         US &     English \\
1017 &                 B &      No &              &                 &                 4 &            Yes &  18 - 24 &                      Male &         US &     English \\
1020 &                HS &     Yes &            U &           Hobby &                 3 &             No &  18 - 24 &                    Female &         US &   Mongolian \\
1022 &                HS &     Yes &            U &    Professional &                 3 &             No &  18 - 24 &                      Male &         US &     English \\
1023 &                HS &     Yes &            U &           Hobby &                 4 &             No &  18 - 24 &                      Male &   Malaysia &     English \\
1024 &                 B &     Yes &            G &    Professional &                 3 &            Yes &  25 - 34 &                      Male &  Indonesia &     Kannada \\
1027 &                HS &     Yes &            U &            None &                 3 &             No &  18 - 24 &                      Male &         US &     English \\
1028 &                HS &     Yes &            U &    Professional &                 4 &             No &  18 - 24 &                    Female &      China &     Chinese \\
1029 &                HS &     Yes &            U &           Hobby &                 3 &             No &  18 - 24 &                      Male &    Myanmar &     Burmese \\
1031 &                HS &     Yes &            U &    Professional &                 4 &             No &  18 - 24 &                      Male &         US &     English \\
1032 &                HS &     Yes &            U &    Professional &                 4 &             No &  18 - 24 &                      Male &         US &     Chinese \\
1033 &                HS &     Yes &            U &           Hobby &                10 &             No &  18 - 24 &                      Male &         US &     English \\
1034 &                HS &     Yes &            U &           Hobby &                 2 &            Yes &  18 - 24 &                      Male &         US &     English \\
1036 &                 A &     Yes &            U &           Hobby &                 3 &             No &  18 - 24 &                    Female &      India &       Hindi \\
1037 &                 B &      No &              &                 &                 7 &            Yes &  18 - 24 &                    Female &         US &     English \\
1038 &                HS &     Yes &            U &            None &                 5 &             No &  18 - 24 &                    Female &      India &     Kannada \\
1040 &                 M &      No &              &                 &                 7 &             No &  18 - 24 &                      Male &      India &             \\
1041 &                 B &     Yes &            U &    Professional &                 8 &            Yes &  18 - 24 &                      Male &         US &     English \\
1042 &                HS &     Yes &            U &                 &                 2 &             No &  18 - 24 &                    Female &         US &       Tamil \\
1043 &                HS &     Yes &            U &           Hobby &                 1 &             No &  18 - 24 &                      Male &      China &     Chinese \\
1045 &                HS &     Yes &            U &            None &                 1 &             No &  18 - 24 &                    Female &      India &       Hindi \\
1046 &                HS &     Yes &            U &    Professional &                 3 &            Yes &  18 - 24 &                    Female &      India &       Hindi \\
2001 &                 B &     Yes &            G &    Professional &                 9 &            Yes &  18 - 24 &                      Male &         US &     Chinese \\
2003 &                 D &     Yes &            G &    Professional &                15 &            Yes &  25 - 34 &                      Male &         US &     English \\
\hline
\end{tabular}
\label{table:experiment_demographics}
\end{table*}

\setcounter{table}{6}
\begin{table*}
\centering
\begin{tabular}{|lllllllllll|}
\hline
\textbf{Control} & education & student & type & experience & years & security &      age &                gender &    country &    language \\
\hline
22   &                HS &     Yes &            U &            None &                 5 &             No &  18 - 24 &                  Male &         US &     English \\
177  &                 B &     Yes &            G &           Hobby &                 3 &            Yes &  18 - 24 &                Female &            &             \\
178  &                HS &     Yes &            U &    Professional &                 7 &             No &  18 - 24 &                  Male &     Brazil &  Portuguese \\
1002 &                 M &     Yes &            G &    Professional &                13 &            Yes &  25 - 34 &                  Male &      China &     Chinese \\
1005 &                HS &     Yes &            U &    Professional &                10 &            Yes &  18 - 24 &                  Male &         US &     English \\
1009 &                HS &     Yes &            U &           Hobby &                 8 &            Yes &  18 - 24 &  Trans Male &         US &     English \\
1012 &                HS &     Yes &            U &           Hobby &                 1 &             No &  18 - 24 &                Female &      China &     Chinese \\
1013 &                HS &     Yes &            U &           Hobby &                 3 &             No &  18 - 24 &                  Male &  Hong Kong &     Chinese \\
1018 &                 B &     Yes &            U &    Professional &                 3 &             No &  18 - 24 &                Female &      China &     Chinese \\
1019 &                HS &     Yes &            U &           Hobby &                13 &             No &  18 - 24 &                  Male &         US &     English \\
1030 &                HS &     Yes &            U &    Professional &                 5 &             No &  18 - 24 &                  Male &         US &     English \\
1035 &                 B &      No &              &                 &                 8 &             No &  18 - 24 &                  Male &         US &     English \\
1039 &                HS &     Yes &            U &    Professional &                 4 &             No &  18 - 24 &                  Male &         US &     English \\
2002 &                 B &     Yes &            G &    Professional &                 7 &             No &  18 - 24 &                  Male &         US &     English \\
\hline
\end{tabular}
\caption{Education contains the highest level of education a participant has achieved, where A is an Associates degree, B is a Bachelors degree, HS is a high school diploma, and D is a Doctoral or Professional Degree. Type contains the type of student, where U is undergraduate and G is graduate. Years contains the number of years of programming experience that a participant has. Security indicates if the participant has taken a security class.}
\label{table:control_demographics}
\end{table*}

\subsection{AI vs non-AI Experiment}
\label{appendix:ainonai}
Table~\ref{table:av_vs_non_ai} shows the attribution of mistakes within the experiment group. While our qualitative coding marks more specific categories, such as ``User+AI+Internet'', for this analysis we bucket all categories that involved the AI Assistant together. 
 
\begin{table}
\centering
\begin{tabular}{|ll|l|l|}
\hline
{} & \textbf{Mistake} & \textbf{AI} & \textbf{non-AI} \\ \hline
Q1 & auth & 58\% & 9\% \\
{} & padding & 12\% & 0\% \\
{} & trivial & 36\% & 6\% \\
{} & mode & 9\% & 0\% \\
{} & library & 0\% & 0\% \\
\hline
Q2 & random & 48\% & 15\% \\
\hline
Q3 & parent & 61\% & 15\% \\
{} & symlink & 73\% & 15\% \\
\hline
Q4 & sql injection & 30\% & 6\% \\
\hline
Q5 & buffer overflow & 12\% & 6\% \\
{} & local pointer & 9\% & 9\% \\
{} & int overflow & 15\% & 3\% \\
\hline
\end{tabular}
\caption{Percentage of mistakes made within the experiment group, broken down by the originator of the mistake (AI vs non-AI).}
\label{table:av_vs_non_ai}
\end{table}

\subsection{Demographics} \label{appendix:demographics}
Table~\ref{table:control_demographics} contains more detailed demographics on the participant population for the experiment and control groups.

\subsection{Regression Tables} \label{appendix:regression}
Table~\ref{table:security_regression} contains the data for the logistic regression used in Section~\ref{security_analysis}.
Data was bucketed as follows. For Q1, ``Secure'' and ``Partially Secure'' answers were grouped as secure. ``Insecure'' answers were grouped as insecure. For Q2, ``Secure'' answers were grouped as secure. ``Partially Secure'' and ``Insecure'' answers were grouped as insecure. For Q3, ``Secure'' and ``Partially Secure'' answers were grouped as secure. ``Insecure'' answers were grouped as insecure. For Q4, ``Secure'' answers were grouped as secure and ``Insecure'' answers were grouped as insecure. For Q5, ``Secure'', ``RC'', and ``DoS'' answers were grouped as secure. ``Partially Secure'' and ``Insecure'' answers were grouped as insecure. ``Partially Secure'' answers were placed into different buckets for different questions due to their varying severity. Note that while this table reports results for the effect of the experiment/control groups,  we determine statistical significance of this treatment for particular security buckets (e.g. only ``Insecure'') using Welch's unequal variance t-test in our main reported results.

\end{document}